\documentclass[12pt,preprint]{aastex}
\def\degpoint{\ifmmode ^{\rm{o}}\!. \else $^{\rm{o}}\!.$\fi}

\newcommand{\ms}{\mbox{m\,s$^{-1}$}}

\newcommand{\Msun}{\mbox{M$_{\odot}$}}

\newcommand{\Mjup}{\mbox{M$_{\rm Jup}$}}

\newcommand{\ltsimeq}{\raisebox{-0.6ex}{$\,\stackrel
         {\raisebox{-.2ex}{$\textstyle <$}}{\sim}\,$}}
\newcommand{\gtsimeq}{\raisebox{-0.6ex}{$\,\stackrel
         {\raisebox{-.2ex}{$\textstyle >$}}{\sim}\,$}}

\begin{document}

\title{On the Frequency of Jupiter Analogs }
\author{Robert A.~Wittenmyer\altaffilmark{1}, 
C.G.~Tinney\altaffilmark{1}, Simon J.~O'Toole\altaffilmark{2}, 
H.R.A.~Jones\altaffilmark{3}, R.P.~Butler\altaffilmark{4}, 
B.D.~Carter\altaffilmark{5}, J.~Bailey\altaffilmark{1} }
\altaffiltext{1}{Department of Astrophysics, School of Physics, 
University of NSW, 2052, Australia}
\altaffiltext{2}{Anglo-Australian Observatory, PO Box 296, Epping, 1710, 
Australia}
\altaffiltext{3}{Centre for Astrophysics Research, University of 
Hertfordshire, College Lane, Hatfield, Herts AL10 9AB, UK}
\altaffiltext{4}{Department of Terrestrial Magnetism, Carnegie 
Institution of Washington, 5241 Broad Branch Road, NW, Washington, DC 
20015-1305, USA}
\altaffiltext{5}{Faculty of Sciences, University of Southern Queensland, 
Toowoomba, Queensland 4350, Australia}
\email{
rob@phys.unsw.edu.au}

\shorttitle{Jupiter Analogs }
\shortauthors{Wittenmyer et al.}

\begin{abstract}

The Anglo-Australian Planet Search has now accumulated 12 years of 
radial-velocity data with long-term instrumental precision better than 3 
\ms.  In this paper, we expand on earlier simulation work, to probe the 
frequency of near-circular, long-period gas-giant planets residing at 
orbital distances of 3-6 AU -- the so-called ``Jupiter analogs.'' We 
present the first comprehensive analysis of the frequency of these 
objects based on radial-velocity data.  We find that 3.3\% of stars in 
our sample host Jupiter analogs; detailed, star-by-star simulations show 
that no more than 37\% of stars host a giant planet between 3--6 AU.

\end{abstract}


\section{Introduction}

Recent discoveries of ever lower-mass planets have received a great deal 
of publicity.  Equally important, however, are discoveries of 
long-period planets.  A long-standing question in astrophysics is ``How 
common are planetary systems like our own Solar system?'' The so-called 
``Jupiter analogs,'' with orbital periods $P\gtsimeq$ 10 years and 
velocity amplitudes $K\sim$ 10 \ms, represent another means for probing 
the frequency of systems with architectures similar to our Solar 
system\footnote{For Jupiter, $P=$11.86yr and $K=$12.5 \ms.}.  Such 
planets are now within the reach of the longest-running radial-velocity 
planet-search programs.  A critical part of addressing the key question 
of Solar system frequency is assessing the selection effects at work in 
the regime of long-period, Jovian planets.


Microlensing observations are beginning to be effective in detecting and 
constraining the population of Jupiter-like planets.  The gravitational 
microlensing technique is well-suited to detecting planets which have 
not experienced significant inward migration, typically at separations 
$a>3$~AU (cf.~Fig.~9 from Sumi et al.~2010).  The discovery of a 
Jupiter/Saturn analog by \citet{gaudi08} illustrates the power of the 
microlensing technique to detect planets which are currently beyond the 
reach (in both observation time and precision) of Doppler measurements.  
\citet{gould10} presented the first estimate of the frequency of planets 
beyond the snow line, based on a sample of 6 microlensing planet 
discoveries.

Previously published analyses of radial-velocity data sets have been 
able to place upper limits on substellar and planetary companions from 
radial-velocity surveys.  \citet{murdoch93} determined detection limits 
for the Mt.~John radial-velocity program by adding the program's mean 
velocity error (65 \ms) to the signals of planets in circular orbits.  
Planetary signals were then recovered by the periodogram and F-test 
methods, and those planets for which 95\% of phases were recovered with 
false-alarm probability (FAP)$<$1\% were considered detectable.  They 
were able to exclude planets with m~sin~$i>$10 \Mjup\ with periods less 
than 2000 days, and brown dwarf companions (10-40 \Mjup) with $P<8.2$ 
years.  Similarly, \citet{cumming99} computed detection limits from the 
Lick planet search data by 1) noting the highest peak $z_{max}$ in the 
periodogram for each target, and 2) generating simulated data sets with 
sinusoidal (circular-orbit) signals and finding the velocity amplitude 
$K$ for which 99\% of signals had power exceeding $z_{max}$.  They 
achieved a detection limit of 20 \ms\ for companions with $a\ltsimeq$5 
AU.  The 12-year CFHT survey of \citet{walker95}, with a velocity 
precision of 15 \ms, achieved detection limits approaching a Jupiter 
mass for planets in circular orbits and periods shorter than 
$\sim$10~yr.  \citet{limitspaper} combined the data of \citet{walker95} 
with data from the McDonald Observatory planet search to achieve a 
baseline of more than 20 years for a sample of 31 bright solar-type 
stars.  Those authors estimated a 99\% detection limit of 2.0$\pm$1.1 
\Mjup\ for planets in Jupiter-like orbits ($e=0.0$, $a$=5.2 AU).  
Recently, \citet{cumming08} presented a detailed analysis of 8 years of 
Keck Planet Search data (585 stars with $N>$10 measurements), resulting 
in a typical detection limit of 10 \ms\ for periods less than the 
duration of the observations.  From those data, \citet{cumming08} 
estimated a giant planet (0.3-10 \Mjup) frequency of 10.5\% for orbital 
periods less than 2000 days.

The Anglo-Australian Planet Search (AAPS) has been in operation since 
1998 January, and currently monitors 250 stars.  The AAPS has achieved a 
long-term radial-velocity precision of 3 \ms\ or better since its 
inception, which is enabling the detection of long-period giant planets.  
These planets have typical velocity amplitudes $K\ltsimeq$15 \ms, and 
their detection requires superb long-term velocity stability.  Some 
notable recent AAPS detections of such planets include HD~134987c (Jones 
et al.~2010: $P=13.7$yr, M~sin~$i$=0.82\Mjup, $K=9$\ms) and GJ~832b 
(Bailey et al.~2009: $P=9.4$yr, M~sin~$i$=0.64\Mjup, $K=15$\ms).  The 
long-term precision of the AAT/UCLES system has enabled the AAPS to be 
relatively efficient at detecting long-period planets.  Of the planets 
discovered by the AAPS, 37$\pm$11\% have periods longer than 1000 days.  
For comparison, this figure is 29$\pm$5\% for the Lick \& Keck program, 
18$\pm$4\% for HARPS+CORALIE+ELODIE, and 14$\pm$6\% for all other 
planet-search teams.  These figures are obtained by counting the number 
of planets with $P>$1000 days discovered by each group, then dividing by 
the total number of planets discovered by that group.\footnote{Planet 
data obtained from the Exoplanet Data Explorer at exoplanets.org} As the 
AAPS now spans 12 years, it is important to make quantitative estimates 
of the population of detectable Jupiter analogs in our sample and in the 
Solar neighbourhood.

Here we define a ``Jupiter analog'' as a planet with a small 
eccentricity ($e<0.2$) and a long period ($P\gtsimeq$8 yr).  That is, a 
giant planet which plays a dynamical role similar to that of our own 
Jupiter, with a period long enough to imply \textit{in situ} formation, 
and an eccentricity low enough to suggest a benign dynamical history.  
In this paper, we present detection limits for Jupiter analogs from the 
12-year AAPS database.  Section~2 briefly describes the observational 
data, and Section~3 discusses the techniques used to calculate detection 
limits.  In Section~4, we present the results, and in Section~5 we 
discuss the results in the broader context of the population of 
long-period planets.

\section{Observations}

AAPS Doppler measurements are made with the UCLES echelle spectrograph 
\citep{diego:90}.  An iodine absorption cell provides wavelength 
calibration from 5000 to 6200\,\AA.  The spectrograph point-spread 
function and wavelength calibration are derived from the iodine 
absorption lines embedded on every pixel of the spectrum by the cell 
\citep{val:95,BuMaWi96}.  The result is a precision Doppler velocity 
estimate for each epoch, along with an internal uncertainty estimate, 
which includes the effects of photon-counting uncertainties, residual 
errors in the spectrograph PSF model, and variation in the underlying 
spectrum between the iodine-free template, and epoch spectra observed 
through the iodine cell.  All velocities are measured relative to the 
arbitrary zero-point defined by the template observation.

The AAPS target list contains 254 stars, of which 180 have been observed 
for more than 8 years.  Since the aim of this work is to place 
meaningful limits on Jupiter analogs, here we only consider those stars 
which have more than 8 years of data ($N=180$).  We further restrict the 
sample to those stars which have been observed more than 30 times 
($N=123$).  This is because the reliability of the FAP calculation is 
strongly dependent on the number of data points.



Table~\ref{rvdata1} summarises the data characteristics for these 123 
stars.  For those stars with long-term trends indicating a distant 
stellar companion, a linear or quadratic fit was removed from the data 
before subjecting them to the detection-limit procedure.  For stars 
known to host a substellar companion, we did a fit for and removed that 
orbit and then performed the detection-limit computations on the 
residuals.

%

\section{Computational Methods}

A set of simulations such as these is only as meaningful as the input 
assumptions and parameters.  In this section, we give a detailed 
discussion of our choices for these simulations.

\subsection{The Detection Limit Algorithm}

The detection limits were computed using the method of 
\citet{limitspaper}.  In brief, we add a Keplerian signal to the 
existing velocity data, then attempt to recover that signal using a 
Lomb-Scargle periodogram.  The mass of the simulated planet is increased 
until 99\% of the injected signals are recovered with FAP$<$0.1\%.  For 
a given mass (or equivalently, a given velocity amplitude) at a given 
orbital period, we use a grid of 30 values of periastron passage $T_0$ 
and, for eccentric orbits, 18 values of the periastron argument $\omega$ 
spaced evenly every 20 degrees.  This makes a total of 30 possible 
orbital configurations for simulated planets with $e=0.0$, and 540 
configurations for those with eccentric orbits.  We simulated planets 
over 100 orbital periods ranging from 1000 to 5000 days, evenly spaced 
in the logarithm of the period.  For a simulated planet with amplitude 
$K$ to be considered detectable at the 99\% level, we must recover 99\% 
of configurations (30/30 for circular orbits, 535/540 for eccentric 
orbits).  A successful recovery occurs when the injected period is the 
highest peak in the periodogram, has a FAP less than 0.1\%, and the 
recovered period is within a specified tolerance of the injected period; 
Section~3.2 gives additional details on the selection of this tolerance.  
In addition to the 99\% recovery level, we performed these simulations 
at recovery levels of 90, 70, and 50\%, to maintain consistency with our 
previous simulation work \citep{monster, foreverpaper}.  Hereafter, we 
refer to each set of simulations by its eccentricity and recovery rate: 
for example, the set where $e=0.1$ at 90\% recovery is referred to as 
``e01r90.''  The result of these simulations is, for each star, a plot of 
the $K$ amplitude (or planet mass) recovered in (99\%, 90\%, 70\%, 50\%) 
of trials at each orbital period between 1000 and 5000 days.

\begin{figure}
\plotone{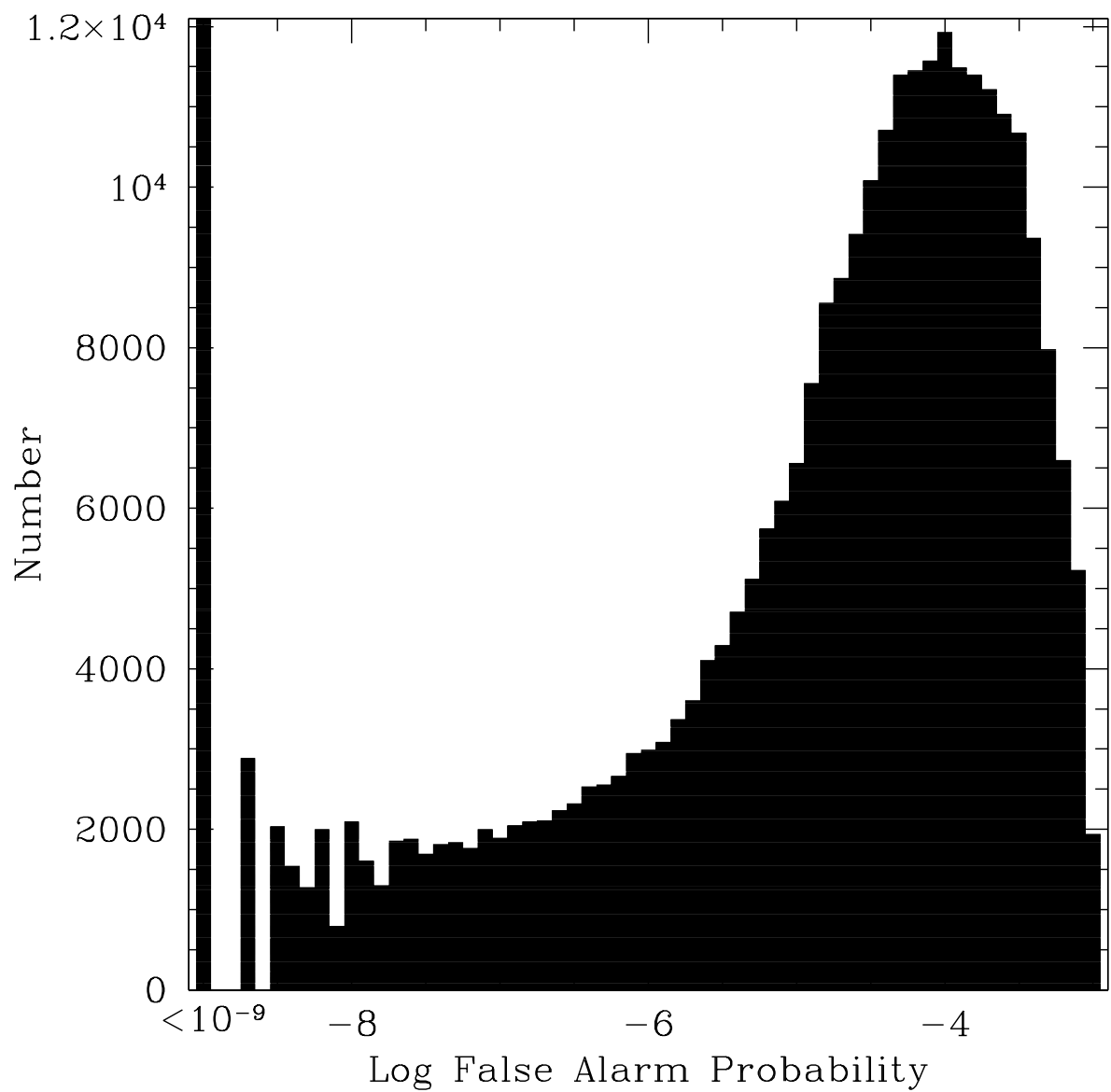}
\caption{Distribution of false-alarm probabilities for 362368 
successfully recovered signals from detection-limit simulations for 123 
stars ($e=0.0$, 99\% recovery). Though the cutoff criterion was 0.001, 
the vast majority of signals were recovered at greater significance, and 
18.4\% of signals were recovered with FAP$<10^{-9}$.  Of all trial 
signals, 23.0\% were rejected based on the FAP cutoff criterion. }
\label{fapsplot}
\end{figure}

Figure~\ref{fapsplot}, which shows the distribution of FAP for all 
simulated signals which were considered successfully recovered in the 
e00r99 trials, demonstrates our reasoning in choosing a cutoff criterion 
of FAP$<$0.1\%.  Summed over all 123 stars, the histogram includes 
results from 362368 simulated signals.  Though the cutoff criterion was 
FAP$<$0.1\% (0.001), it is evident that the vast majority of recovered 
signals achieved FAP values that are far more significant.  The 
distribution has two peaks, at $10^{-4}$ ($N\sim$12000) and $<10^{-9}$ 
($N=66784$).  The latter of these is an artifact, simply representing 
the integration over a long tail of FAP$<10^{-9}$.  If most trials 
resulted in FAP$>10^{-3}$, the cutoff we have imposed would bias the 
derived detection limits.  We would see such an effect in 
Figure~\ref{fapsplot} as a pile-up at the least significant bin 
(FAP$=10^{-3}$).  That the highest peaks in Figure~\ref{fapsplot} lie at 
FAP levels at least an order of magnitude more significant than the 
imposed cutoff indicates that our choice of cutoff has not biased the 
results.

\begin{figure}
\plotone{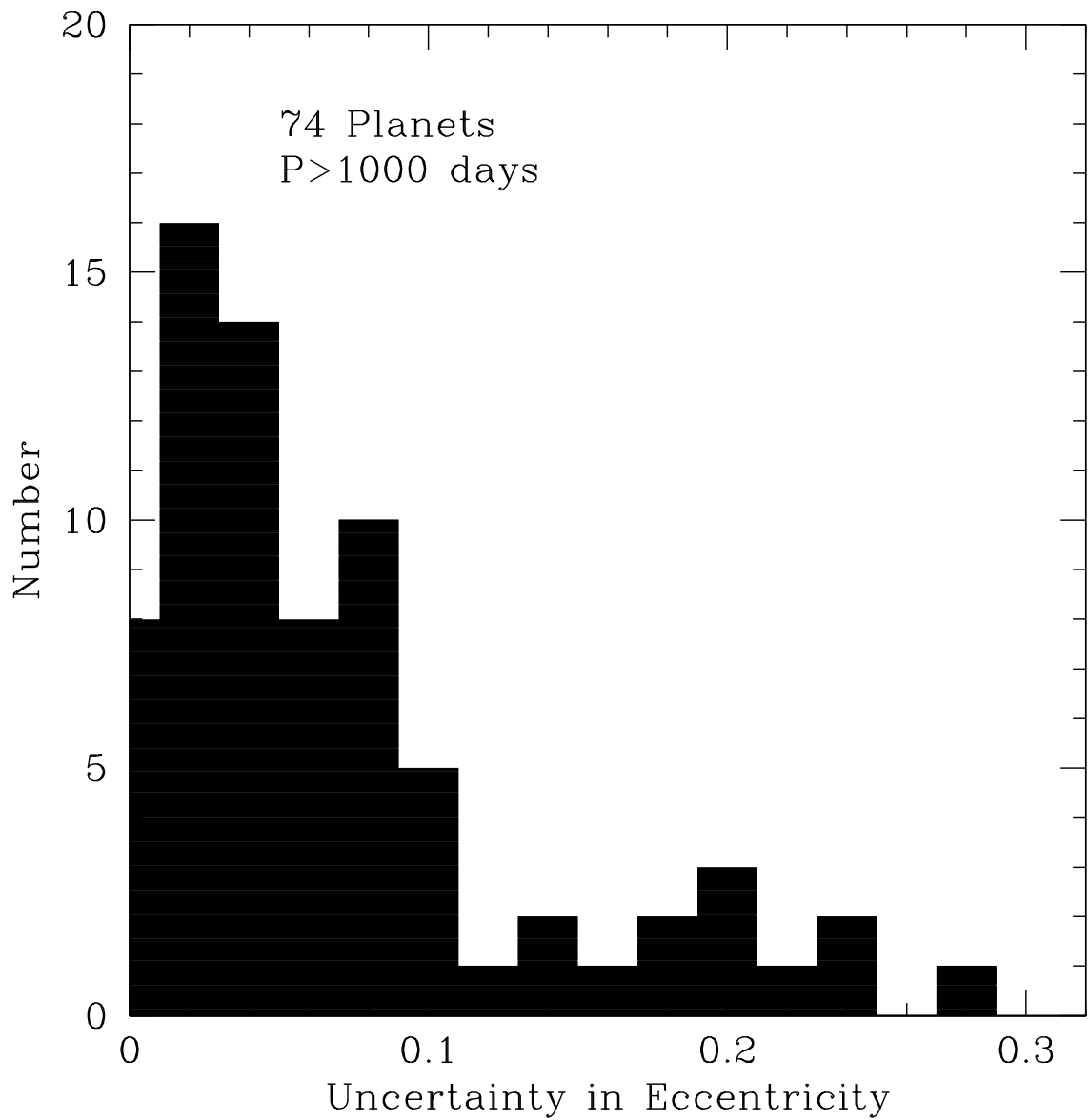}
\caption{Distribution of the published uncertainties in eccentricity for 
the 74 published long-period planets ($P>$1000 days) based on 
exoplanet.eu, 2010 September 1.  The distribution peaks in the range 
$\sigma_e$=0.02-0.04; the true uncertainty in eccentricity may be 
considerably larger due to non-Gaussianities as described in 
\citet{ford05} and \citet{otoole09a}. }
\label{eccentricities}
\end{figure}

In order to focus on potential Jupiter analogs, we only consider 
simulated planets with $e=0.0$, 0.1, and 0.2, in keeping with the 
$e<0.2$ definition of ``Jupiter analog'' described in Section~1.  These 
three values of $e$ provide sufficient sampling of the relevant 
eccentricity range because, for long-period planets ($P>$1000 days), the 
orbital eccentricity is typically determined only to an accuracy of 
0.02-0.04.  Figure~\ref{eccentricities} shows a histogram of the 
uncertainty in eccentricity ($\sigma_e$) for the 74 published planets 
with $P>$1000 days.  The peak of the distribution lies at 
$\sigma_e~\sim$ 0.02-0.04, with a median value of 0.05.  We also note 
that published estimates of $\sigma_e$ are often underestimated due to 
correlations between the Keplerian orbital parameters and the 
non-Gaussianity of their distributions \citep{ford05, otoole09a}.  In 
particular, \citet{otoole09a} found that the 99\% confidence interval 
for eccentricity can be 10-50 times larger than the traditional 
uncertainty estimates derived from the covariance matrix in a 
least-squares Keplerian fit.  This is particularly important when the 
signal of the planet (defined in O'Toole et al.~2009a as $K/\sigma_K$) 
is smaller than about 3.  For these reasons, we have chosen a coarse 
grid of eccentricities for these simulations.  A finer interval would 
vastly increase the computing time required without adding meaningful 
information.

\subsection{False Positives and False Negatives}

All previous implementations of this detection-limit algorithm 
\citep{limitspaper, wittenmyer07, wittenmyer09, foreverpaper} have used 
the criterion that the recovered period be within 5\% of the input 
period of the simulated planet.  The reason for this seemingly arbitrary 
criterion is that in the Lomb-Scargle periodogram, spurious peaks can 
arise due to the sampling of the data.  Most common in radial-velocity 
data are the 1-year and 1-month aliases, the former due to targets 
becoming unobservable as they pass behind the Sun, and the latter due to 
telescope scheduling constraints which usually restrict planet-search 
observations to bright lunations.  Harmonics of the injected 
periodicities (e.g.~$P/2$,$2\times P$, etc.) also often produce 
significant periodogram peaks.  Imposing the ``correct-period'' 
criterion reduces the effect of these features and thus minimises 
incorrect detections (false positives).  For simulations such as these, 
involving a vast number of attempted planetary detections and using 
automated detection criteria, it is critical to understand and to 
minimise (or ideally, to eliminate) false positives.  \citet{otoole09a} 
presented a detailed discussion of the problem of false positives in 
simulations of planet-search data.

\begin{figure}
\plotone{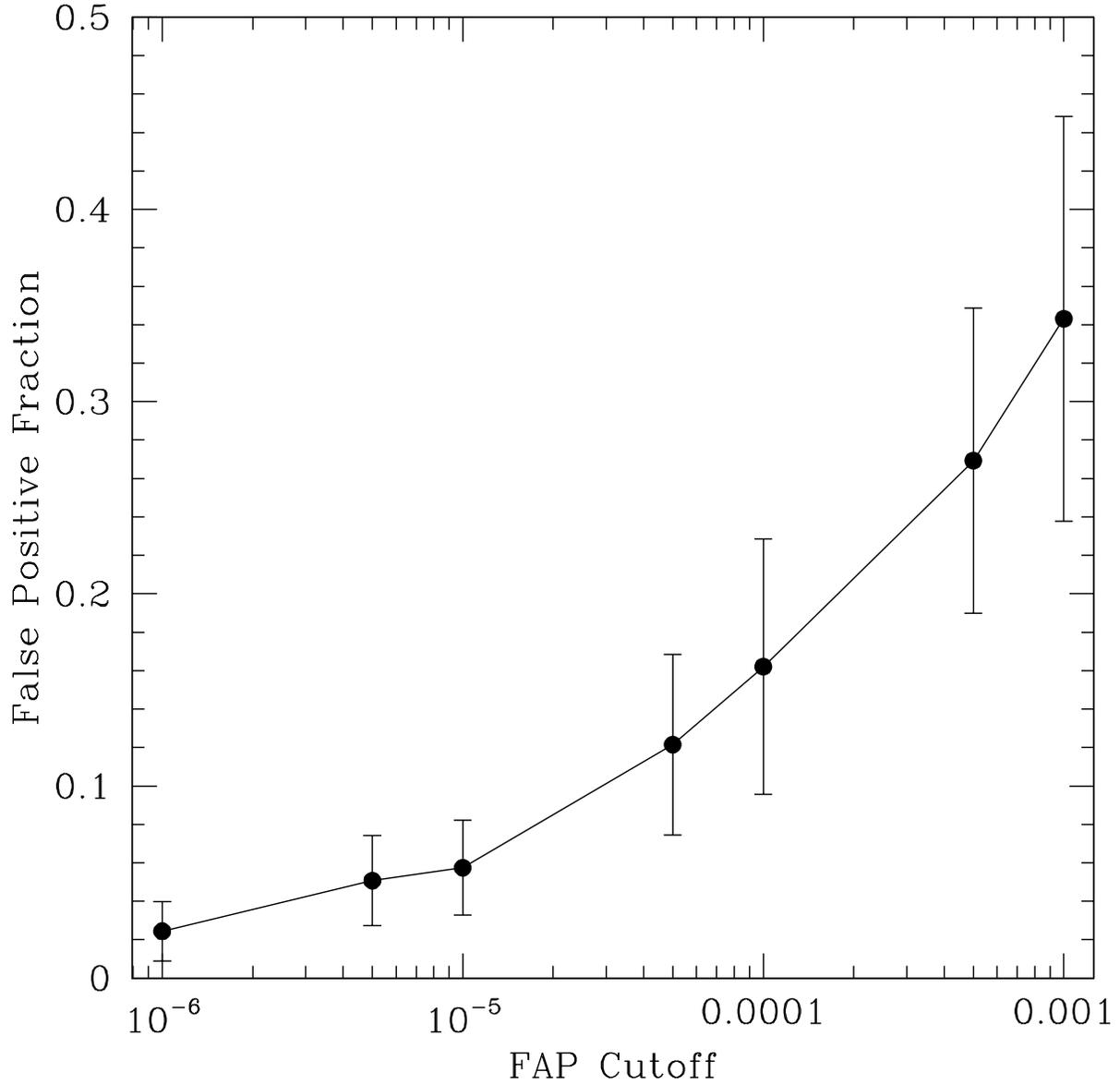}
\caption{Mean false positive rate obtained from tests of the
detection-limit routine in which the correct-period criterion was 
removed. Incorrectly recovered periods tend to have higher FAPs, and 
should be excluded, but a nontrivial number of false positives occur 
even at extremely stringent FAP levels. }
\label{blind}
\end{figure}

A potential pitfall in establishing detection criteria to minimise false 
positives is that, if those criteria are too stringent, false 
\textit{negatives} become important.  False negatives (i.e.~incorrect 
rejections) are substantially more difficult to quantify, and hence are 
fiendishly difficult to control.  In practice, because scientists 
rightfully consider incorrect detections to be a far more serious 
problem, much more effort is spent in eliminating false positives than 
false negatives.  In the detection-limit algorithm described above, we 
seek to eliminate false positives by the imposition of the 
``correct-period'' criterion.  The FAP criterion alone (FAP$<$0.001) 
eliminates some incorrect detections, but since the FAP calculation is 
dependent on the number of data points, even alias periodicities achieve 
high significance for larger data sets ($N\gtsimeq$50).  We performed 
some tests with AAPS data (which have $N>100$ for many targets), 
removing the correct-period criterion and relying on only the FAP 
criterion to eliminate false detections.  Figure~\ref{blind} shows the 
effect of different settings for the FAP criterion in this test.  As 
expected, requiring a more significant periodogram recovery (i.e.~a 
lower FAP cutoff) reduces the false-positive rate.  However, there are 
two important consequences: 1) The false-positive rate remains at a 
nontrivial level even for extremely stringent FAP levels, and 2) When 
the FAP cutoff is set to such low levels, the detection-limit result is 
dominated by false negatives: nearly all of the injected trial signals 
are rejected, and no meaningful detection limit is obtained.

\begin{figure}
\plotone{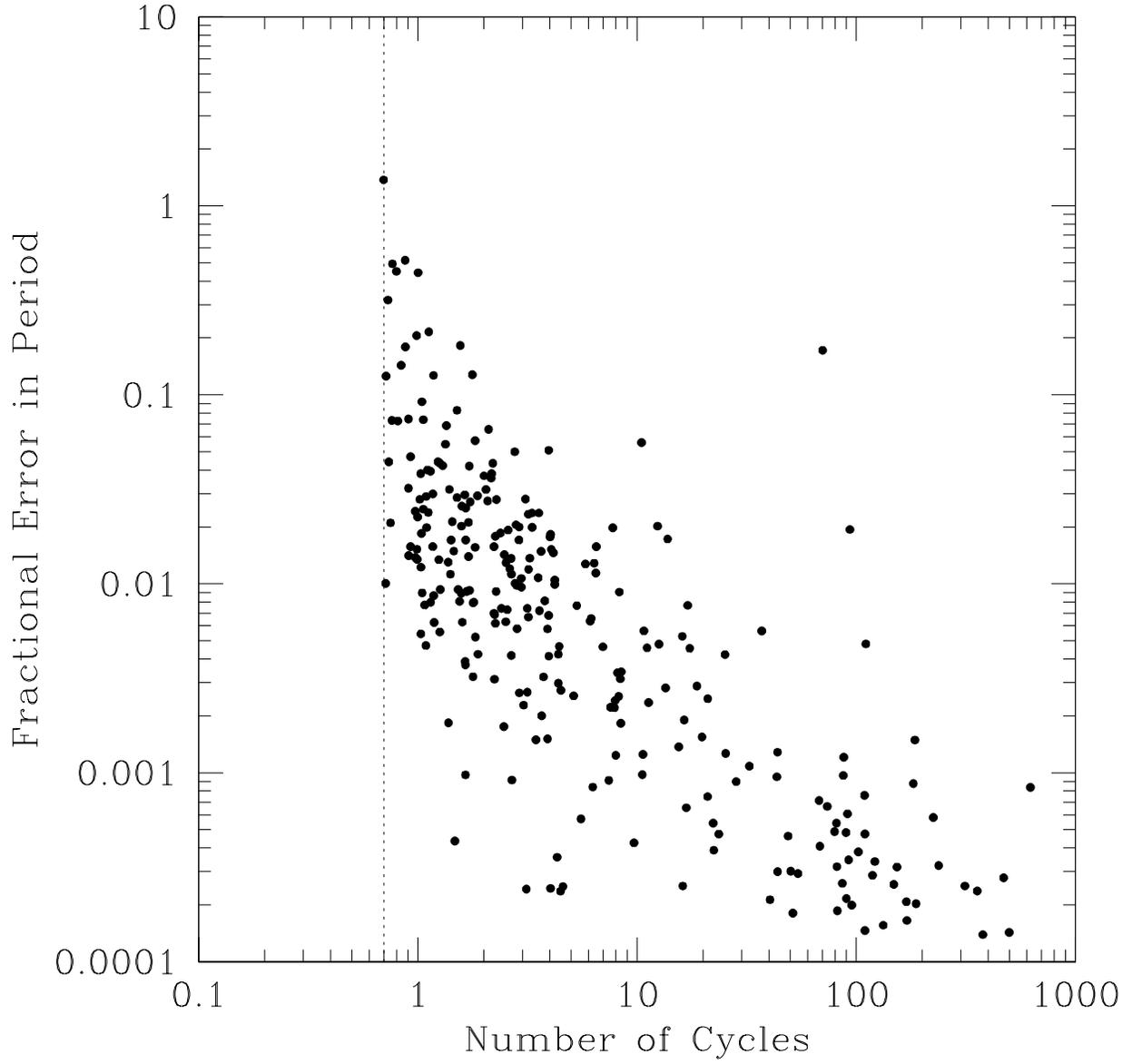}
\caption{Fractional error in orbital period as a function of the number 
of cycles observed, for 290 published radial-velocity planets.  The vertical 
dashed line indicates 0.7 cycles of data, which appears to be the 
minimum for publication. }
\label{published}
\end{figure}

These tests demonstrated that a correct-period constraint is necessary 
to obtain useful detection limits and to control the false positive 
rate.  The next question is then: How close to the input period does the 
recovered period need to be, to be considered a successful detection?  
Too loose a constraint would degrade the scientific value of the 
detection limits obtained from this method, whereas too stringent a 
constraint results in the automated rejection of signals which a 
rational human investigator would consider sufficiently close 
(e.g.~P=3.05 versus 3.04 days).  The core question is: ``What would a 
human do?'' One way to approach this problem is to determine what other 
humans have already done.  That is, what is the fractional uncertainty 
in orbital period at which authors have decided to publish a planet 
detection?  Radial-velocity planet search teams generally prefer to wait 
until a complete orbital cycle has been observed before publishing a 
planet discovery.  One expects the uncertainty in the fitted period 
($\sigma_P$) to drop as more orbital cycles are observed.  We performed 
an exhaustive literature search, locating the original discovery paper 
for every radial-velocity planet, and tabulating the period ($P$), 
period uncertainty ($\sigma_P$), and the total time-span of the 
observations ($\Delta~T$).  In this search, we included only those 
planets: 1) which were originally detected by the radial-velocity 
method, and 2) where a refereed paper is available indicating the 
above-mentioned quantities.  The reason for excluding planets discovered 
by transits is that the transit is a special circumstance which enables 
the orbital period to be determined to extremely high precision.  That 
would skew the relation between $\sigma_{P}/P$ and $\Delta~T$, which we 
seek to apply to the radial-velocity database of the Anglo-Australian 
Planet Search (which includes no transiting planets).  The results of 
this search are shown in Figure~\ref{published}.  One readily evident 
feature is that no planets have been published with less than 0.7 cycles 
of data (vertical dashed line in the figure).  Of the 290 planets, only 
24 (8.3\%) were published with less than one orbital cycle of data.  The 
outlier with an unusually large fractional error in period 
($\sigma_{P}/P$=17.2\%) is HD~149143 \citep{fischer06}.  That planet was 
first published with only 17 observations, which may account for the 
large uncertainty in its 4-day period.

\begin{figure}
\plotone{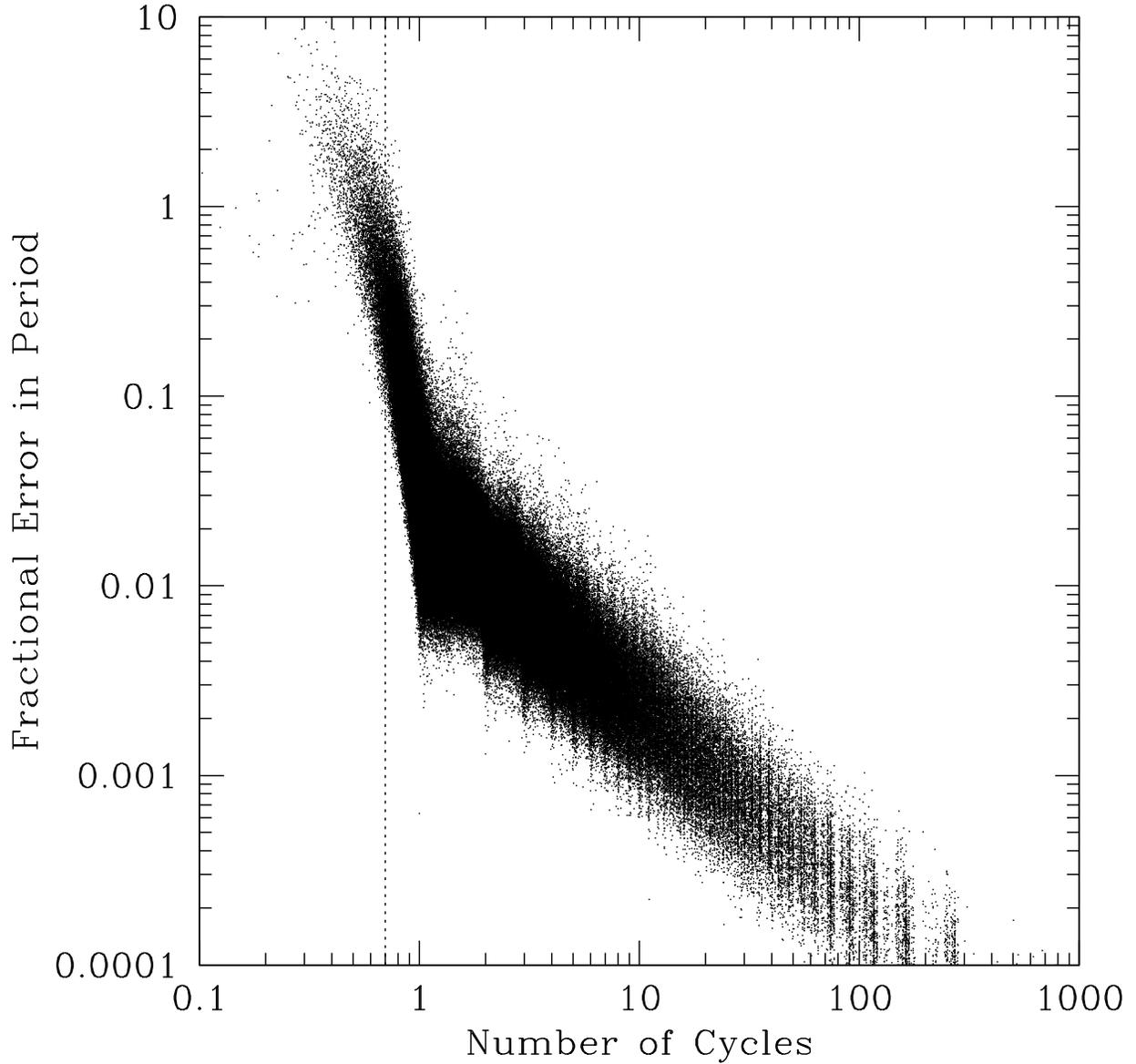}
\caption{Fractional error in orbital period as a function of the number 
of cycles observed, for 362953 simulated radial-velocity planets.  As in 
Figure~\ref{published}, the vertical dashed line indicates 0.7 cycles of 
data. Note that the slope of the relation becomes markedly steeper when 
less than one cycle is available.  }
\label{simulated}
\end{figure}

We wish to derive a relation between the data span $\Delta~T$ 
(equivalently, the number of cycles), and the period error 
$\sigma_{P}/P$; however, the number statistics of the known planets 
remain poor.  Hence, we performed some additional simulations to fill 
out the $\Delta~T$-$\sigma_{P}/P$ plane.  For each of 210 stars in the 
AAPS database, we created 2000 simulated radial-velocity datasets.  Each 
simulated dataset used the observation times and velocity uncertainties 
of the original observed data.  The simulated planets had periods 
ranging from 5 to 5000 days (in a grid with a uniform spacing of 10 
days), velocity amplitudes between 20-50 \ms, and eccentricities 
randomly drawn from the range ${0.0:0.5}$.  Noise (jitter) drawn from a 
Gaussian distribution with a width $\sigma~=5$ \ms\ was also added to 
each simulated velocity measurement.  We then fit each dataset with a 
Keplerian model using GaussFit \citep{jefferys87} and recorded the 
fitted period and its uncertainty.  Since the maximum period used 
sometimes significantly exceeded the length of the available data for 
certain of the targets in the AAPS database, those fits often produced 
unphysical results, and were discarded.  Consistent with the 
characteristics of published planets described above, we also discarded 
all fits where less than 0.7 cycles of data were present.  Of the 420000 
simulations, 357681 remained after these cuts, and the results are 
plotted in Figure~\ref{simulated}.  We can now use these results to make 
an informed choice for the correct-period constraint.  Since the 
injected trial period is assumed not to be known \textit{a priori}, it 
is logical to choose the value of $\sigma_{P}/P$ which includes 99\% of 
the planets simulated in this section.  This results in 
$\sigma_{P}/P=27.7\%$, which is the value used for the correct-period 
constraint in all detection-limit simulations here.

\begin{figure}
\plottwo{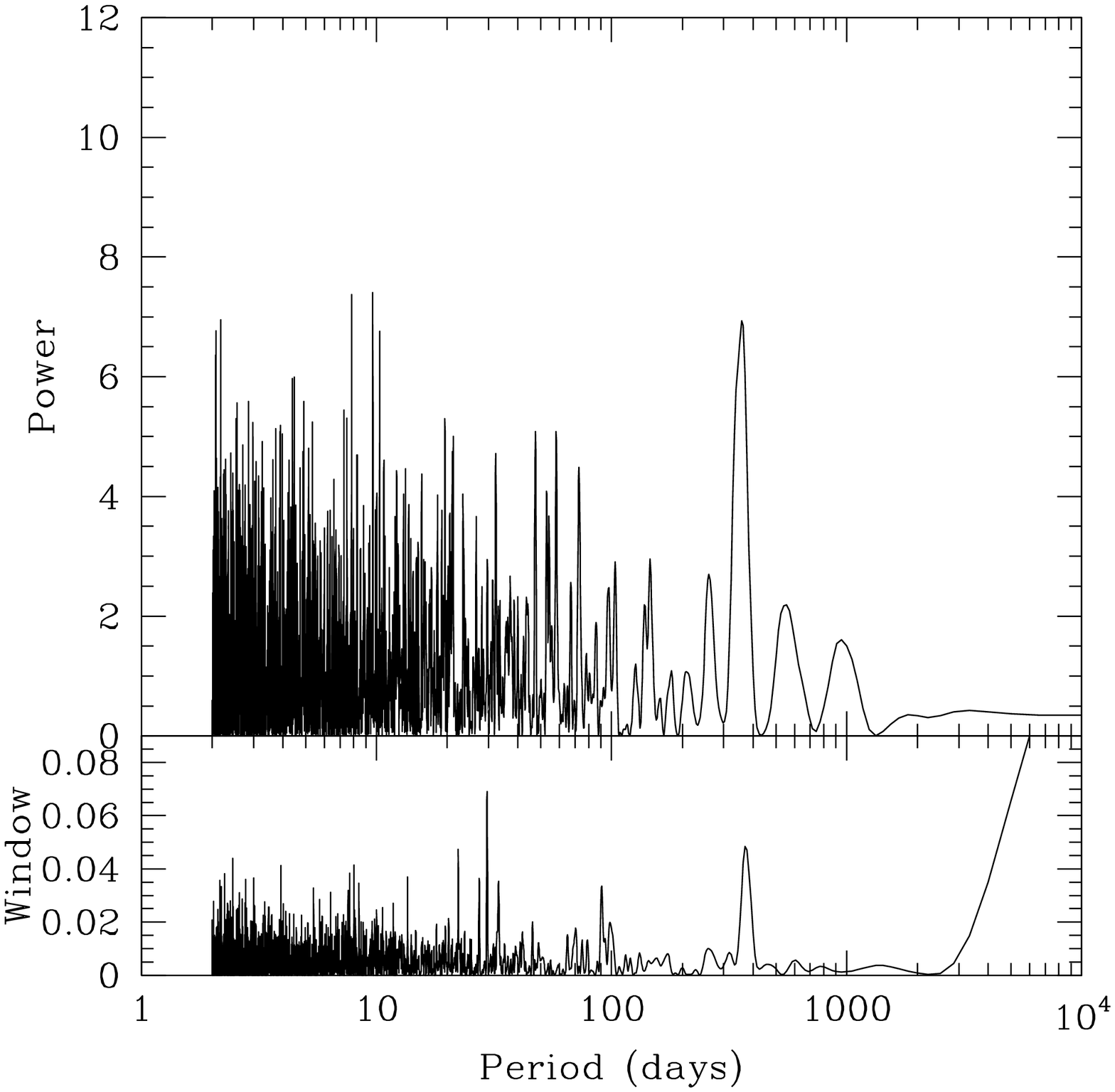}{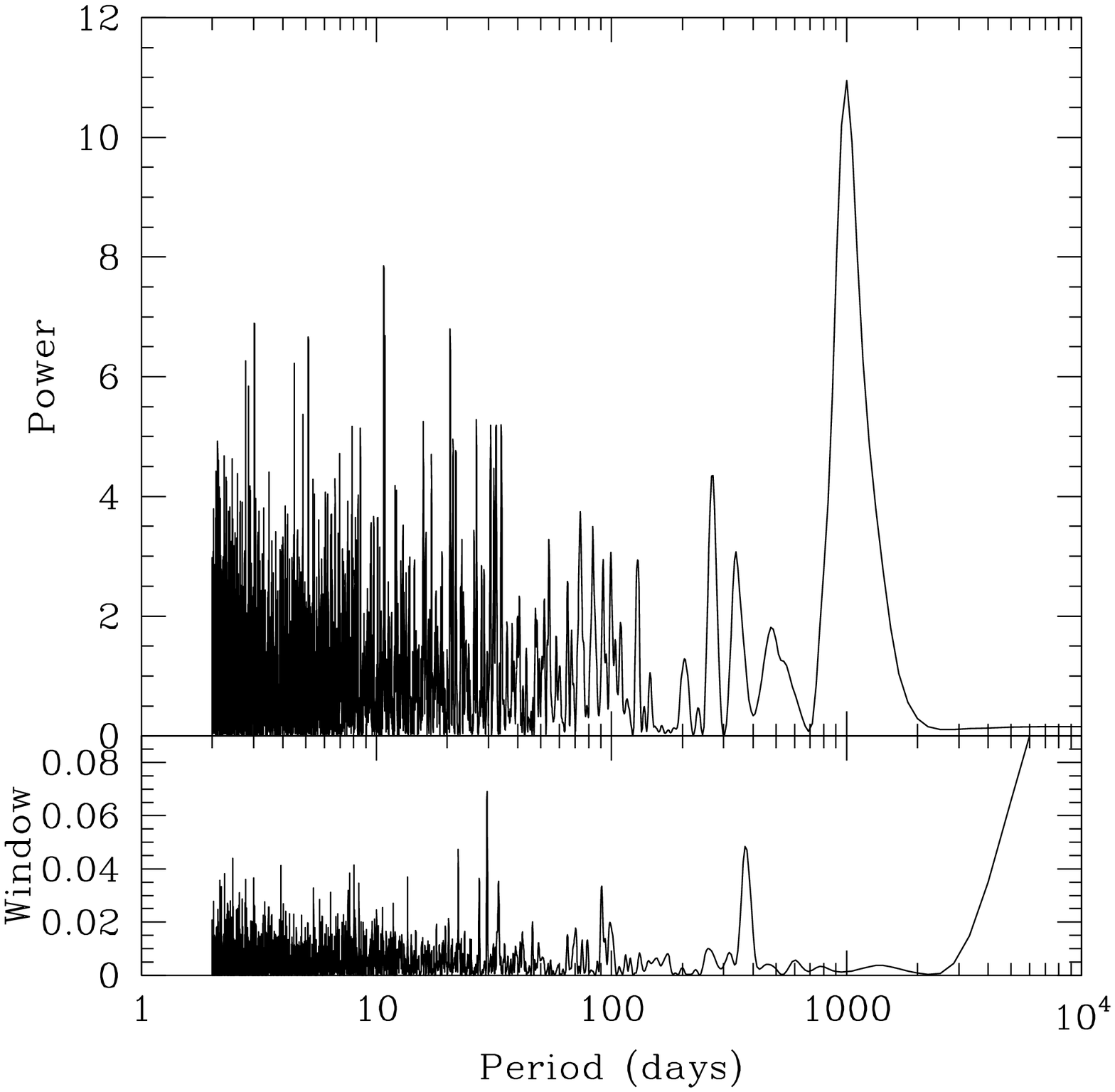}
\caption{An example of a rejected (left panel) and an accepted (right 
panel) detection using the method and criteria outlined in Section~3.  
Both periodograms result from the addition of a signal at $P=1000$ days 
to the velocity data for HD~209653.  The injected signals have $K=$ 
1 \ms\ (left) and 10 \ms\ (right). }
\label{examples}
\end{figure}

Finally, we show in Figure~\ref{examples} examples of a rejected and an 
accepted recovery using the methods described in this section.  An input 
data file (HD~209653) was chosen at random for this demonstration.  The 
left panel shows the periodogram resulting from an injected signal with 
$P=$1000 days, $K=$ 1\ms, and $e=$0.0.  As expected with such a weak 
signal, there is no compelling evidence for it in the periodogram; the 
highest peaks occur at 9.7 and 365 days, the latter an artefact of the 
window function (lower panels).  The FAP of the 9.7-day peak was 0.018, 
much less significant than the cutoff of 0.001.  The right panel of 
Figure~\ref{examples} shows the periodogram resulting from the same 
dataset with an injected signal having $P=$1000 days, $K=$ 10\ms, and 
$e=$0.0.  The peak at 1000 days is quite obvious, and its FAP is 0.0005, 
resulting in a successful recovery.

\section{Results and Discussion}

\begin{figure}
\plotone{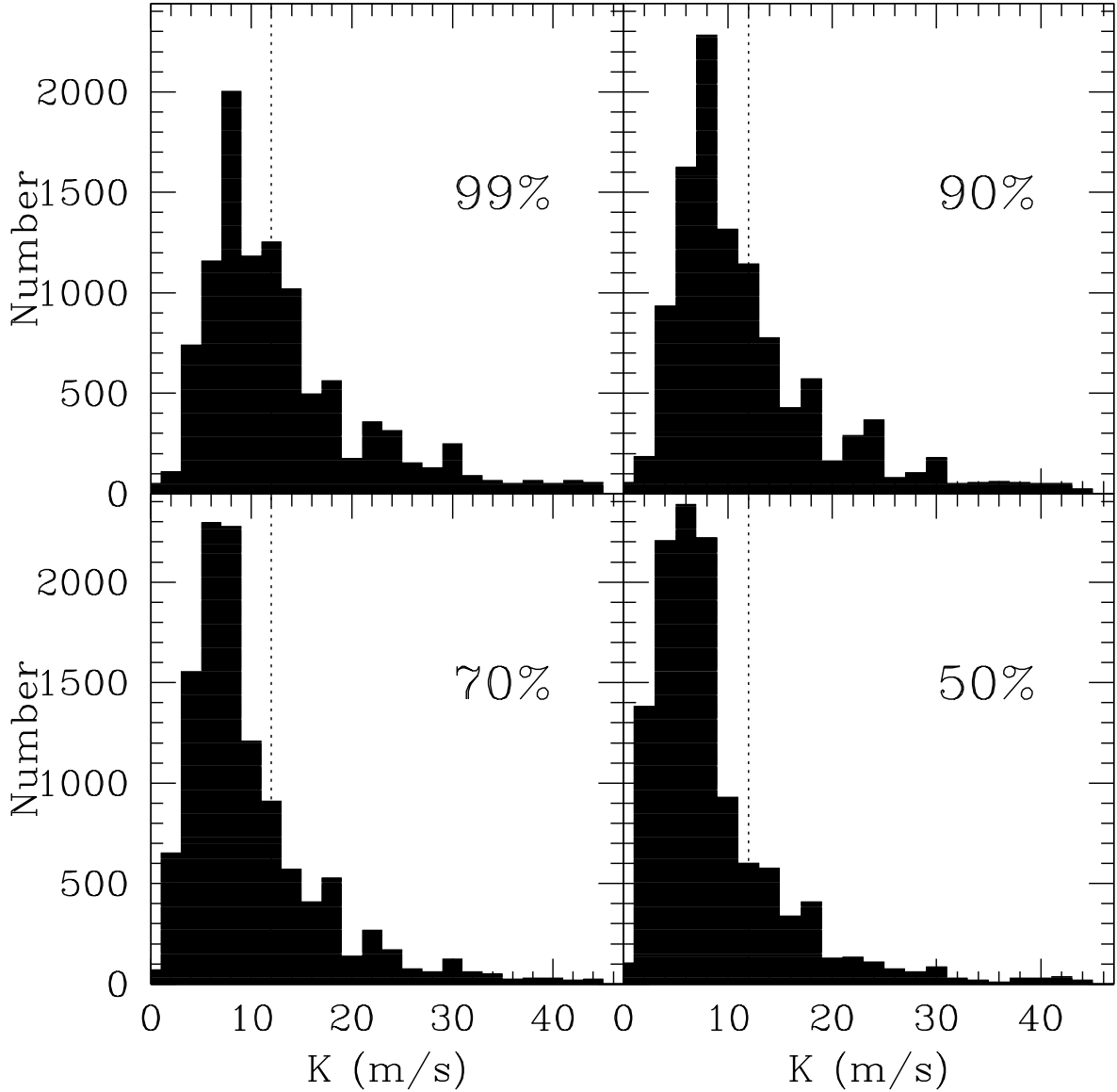}
\caption{Distribution of velocity amplitudes ($K$) recovered for 
simulated planets with $e=0.0$, at recovery levels of 99, 90, 70, and 
50\%.  Each panel shows results from all 123 stars (12300 trial $K$ 
values).  Only $K<50$\ms\ is shown; this range includes 91.7\% of 
trials.  The vertical dotted line at $K=12$ \ms\ indicates the 
radial-velocity signal of Jupiter. }
\label{Khisto1}
\end{figure}

\begin{figure}
\plotone{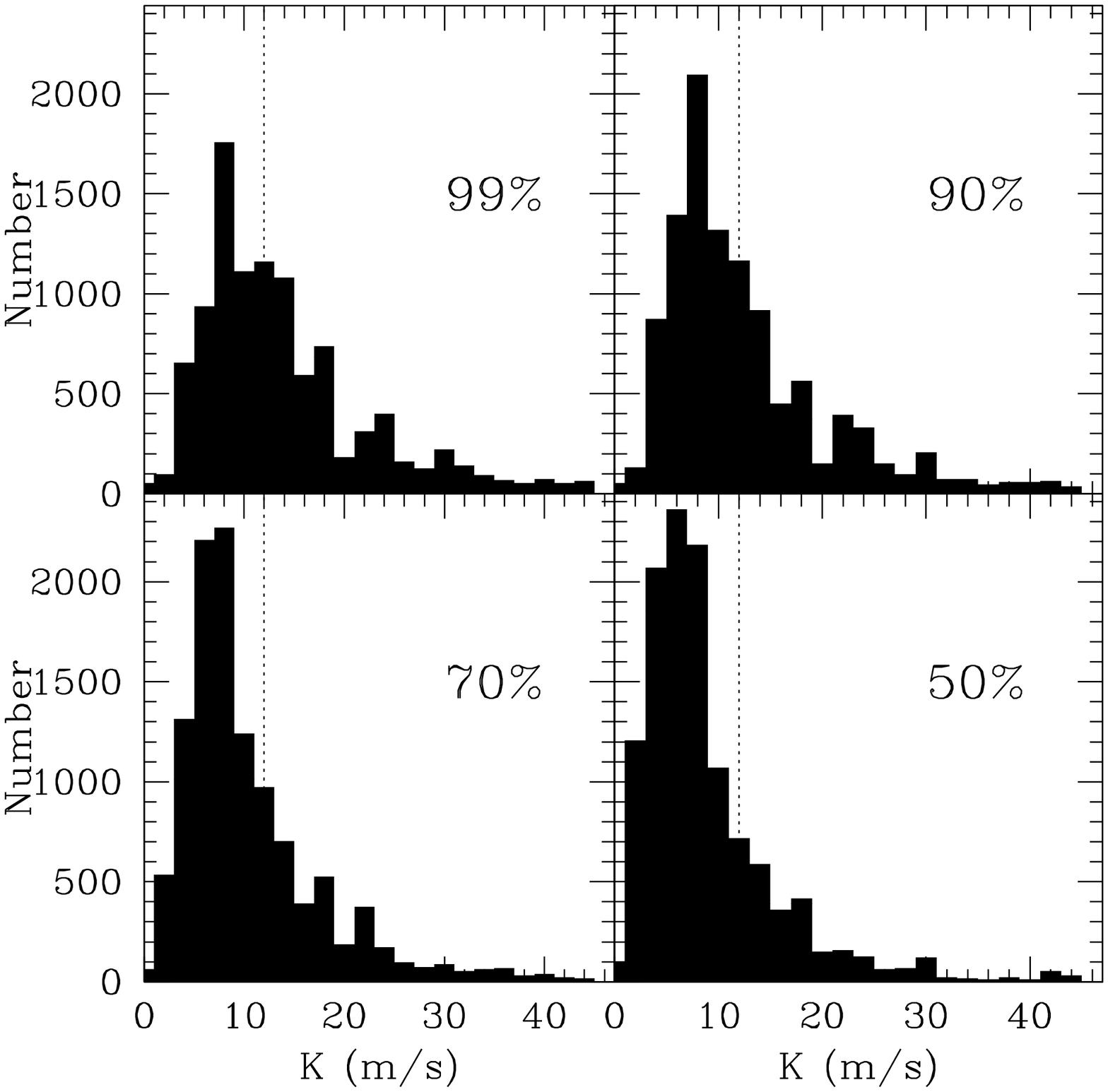}
\caption{Same as Figure~\ref{Khisto1}, but for the $e=0.1$ results.  
Only $K<50$ \ms\ is shown; this range includes 90.4\% of trials. }
\label{Khisto2}
\end{figure}

\begin{figure}
\plotone{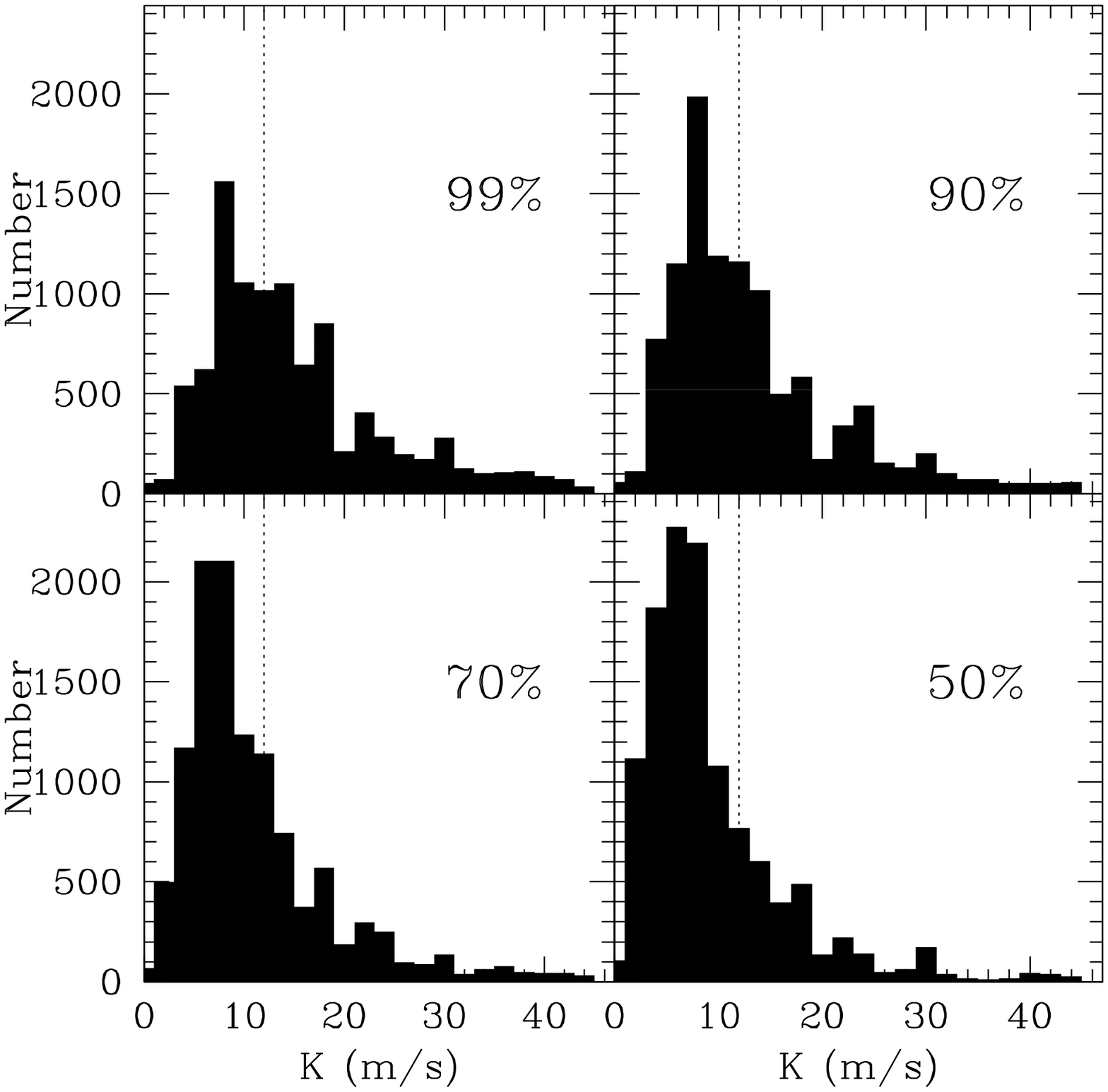}
\caption{Same as Figure~\ref{Khisto1}, but for the $e=0.2$ results.  
Only $K<50$ \ms\ is shown; this range includes 88.8\% of trials. }
\label{Khisto3}
\end{figure}

Due to the large number of stars considered here, it is most efficient 
to present the detection-limit results in terms of the radial-velocity 
amplitude $K$ which the simulations indicate was detectable.  Since the 
targets in the AAPS long-term program are typically solar-type stars, 
one can then estimate the mass thresholds by assuming a 1\Msun\ star.  
Figure~\ref{Khisto1} shows the distribution of $K$ resulting from the 
$e=0.0$ simulations, summed over all stars (123 stars; 12300 $K$ 
values).  As noted in previous work using this detection-limit algorithm 
\citep{limitspaper, wittenmyer09, foreverpaper}, some injected trial 
periods can result in apparently undetectable signals, due to poor time 
sampling or large velocity scatter in the input data.  Circular-orbit 
results are shown in Figure~\ref{Khisto1}, $e=0.1$ results are shown in 
Figure~\ref{Khisto2}, and $e=0.2$ results are shown in 
Figure~\ref{Khisto3}.

\begin{figure}
\plotone{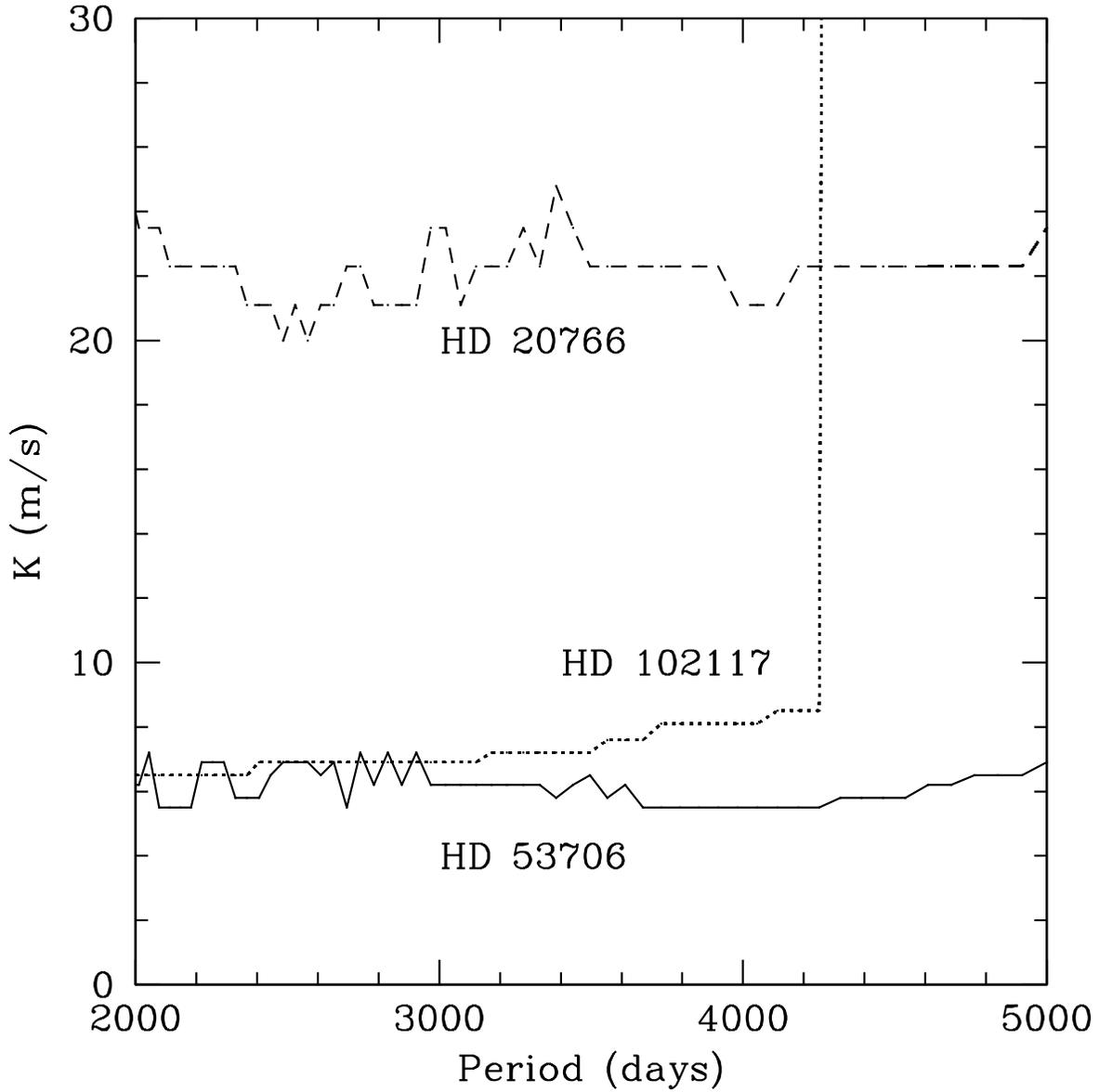}
\caption{Detection limits computed for three representative stars, at 
99\% recovery.  In terms of the velocity amplitude $K$, the detection 
limit is essentially constant for a given star, for periods shorter than 
the duration of observations.  At periods longer than the observational 
data, the injected test signals cannot be robustly detected by our 
automated criteria, as seen for HD~102117 here ($T_{obs}$=3961 days). }
\label{Kexample}
\end{figure}

For a given star, the detection limit in $K$ is generally constant over 
the entire range of trial periods shorter than the duration of 
observations.  Figure~\ref{Kexample} shows the detection limit in $K$ 
for three representative stars.  It is useful, then, to compute the mean 
value of $K$ as the metric of the quality of a star's detection limit.  
For each star, we compute the mean $K$ ($\bar{K}$) and its scatter 
$\sigma_K$, then exclude any values more than 3$\sigma_K$ away from that 
mean and recalculate the mean and its uncertainty.  Summed over all 
stars and all trials, there are 125,100 periods which are shorter than 
the duration of observations, and a total of 110 $K$ values which were 
rejected by this 3$\sigma_K$ clipping process.  The discrepant $K$ 
values tend to occur at longer periods, especially those longer than the 
observations.  The results are given in Table~\ref{meankays} for all 
eccentricities and recovery rates.  There is a wide range of $\bar{K}$, 
spanning two orders of magnitude.  This highlights the importance of 
considering the detection limits on a star-by-star basis.  For example, 
HD~19632, the star for which we obtained the worst detection limits 
($\bar{K}>100$\ms), has a chromospheric activity index log $R'_{HK}$ 
\citep{noyes84} of $-4.38$.  This is one of the most active stars on the 
AAPS target list; ``quiet'' planet-search target stars typically have 
log $R'_{HK}\sim -5.0$.  Radial-velocity modulation due to starspots can 
produce such high levels of jitter \citep{paulson04, wright05}, making 
the determination of meaningful limits extremely difficult.  The 
velocity scatter for the 33 data points on HD~19632 is 26.2\ms, and the 
data show a strong periodicity near 6 days, consistent with its 
estimated rotation period of 12 days (G.~Henry, personal communication).

\begin{figure}
\plotone{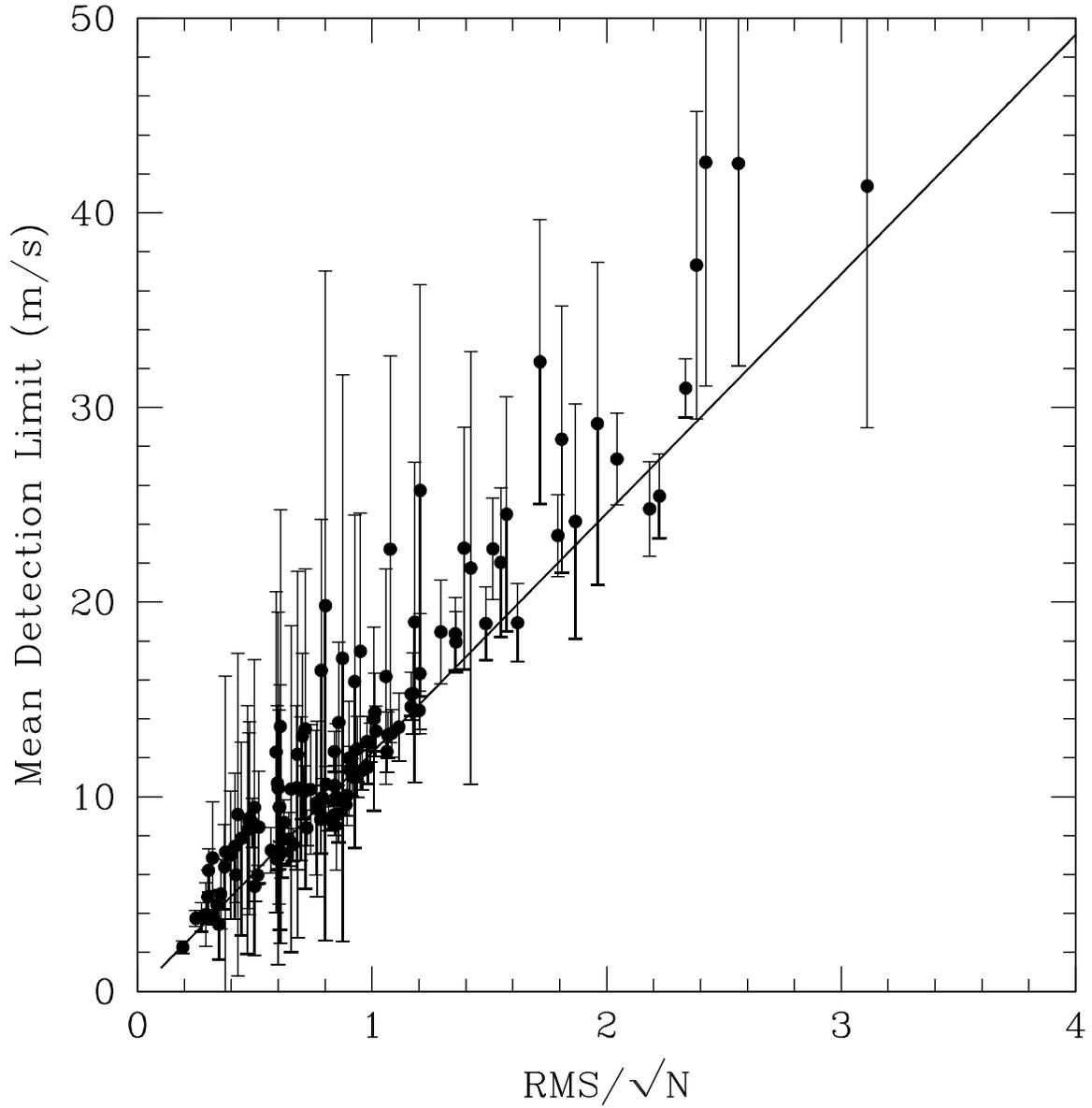}
\caption{Radial-velocity amplitude $K$ that can be detected at the 99\% 
confidence level, plotted against $RMS/\sqrt(N)$, for 119 stars.  A 
linear relation can be fit, with a scatter of 3.6 \ms.  The solid line 
is a weighted fit, which is dominated by the abundance of points at 
$K<15$ \ms\ with small error bars. }
\label{relation}
\end{figure}

The detection limit achievable from a set of radial-velocity data 
depends largely on two factors: the number of data points, and their RMS 
velocity scatter about the mean.  Using these results, we can determine 
an empirical relation between these quantities and the radial-velocity 
amplitude $K$ which can be detected.  Such a relation can then be used 
to estimate the amount of observing time required to obtain a robust 
detection (or non-detection) of a particular class of planet.  
Figure~\ref{relation} plots the mean $K$ obtained for each star 
($e=0.0$, 99\% recovery) versus the quantity $RMS/\sqrt{N}$.  The 119 
stars for which $\bar{K}<50$ \ms\ were used in a linear fit, yielding 
the following relation:

\begin{equation}
K = -0.02 + 12.3\Big{(}\frac{RMS}{\sqrt{N}}\Big{)} \hspace{3mm}\ms.
\end{equation}

Using this fit, we can make estimates of the number of additional 
observations required to place robust constraints on planets in the AAPS 
program.  For example, a star with 40 observations at a total velocity 
rms of 3~\ms\ would yield a detection limit of 5.8~\ms, corresponding to 
0.5\Mjup\ ($P=12$ yr), and 0.1\Mjup\ ($P=50$ d).  It is important to 
note that the cadence of the observations can also have an impact on the 
types of planets which are or are not detectable.  In particular, 
high-cadence observations are more effective at detecting short-period, 
low-amplitude planets, as discussed further in the next subsection.  We 
note that recent modifications to the AAPS observing strategy, such as 
20-minute integrations to average over stellar oscillation noise, have 
resulted in velocity rms of 1-2 \ms\ since 2005 for solar-type stars 
\citep{16417paper, 61vir, foreverpaper}.  The AAPS plans to obtain a 
further 6 years of data applying these strategies, which should markedly 
improve the velocity rms achieved (cf.~Table~\ref{rvdata1}), directly 
resulting in even tighter constraints on Jupiter analogs.

\subsection{Known planet hosts }

\begin{figure}
\plottwo{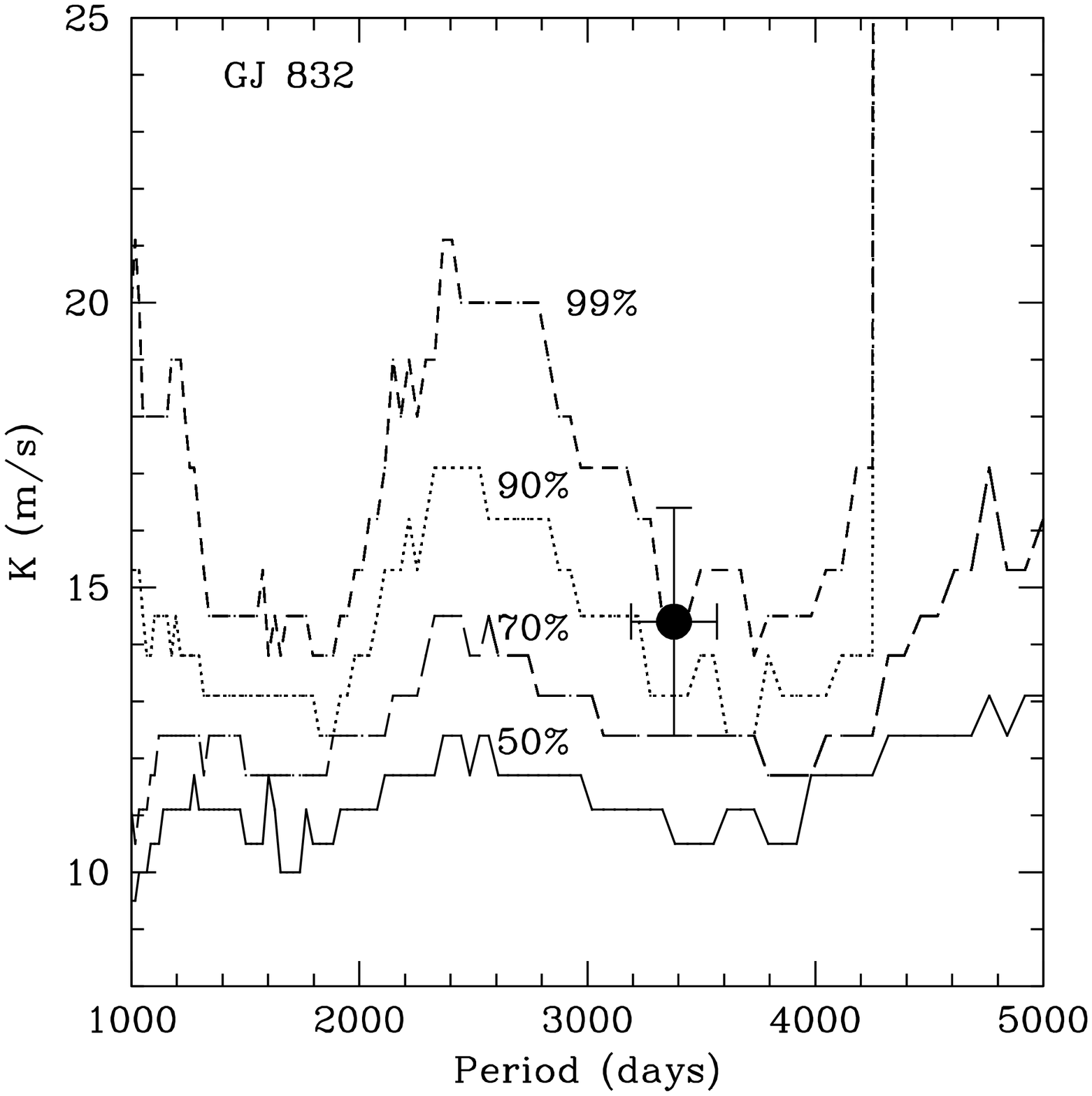}{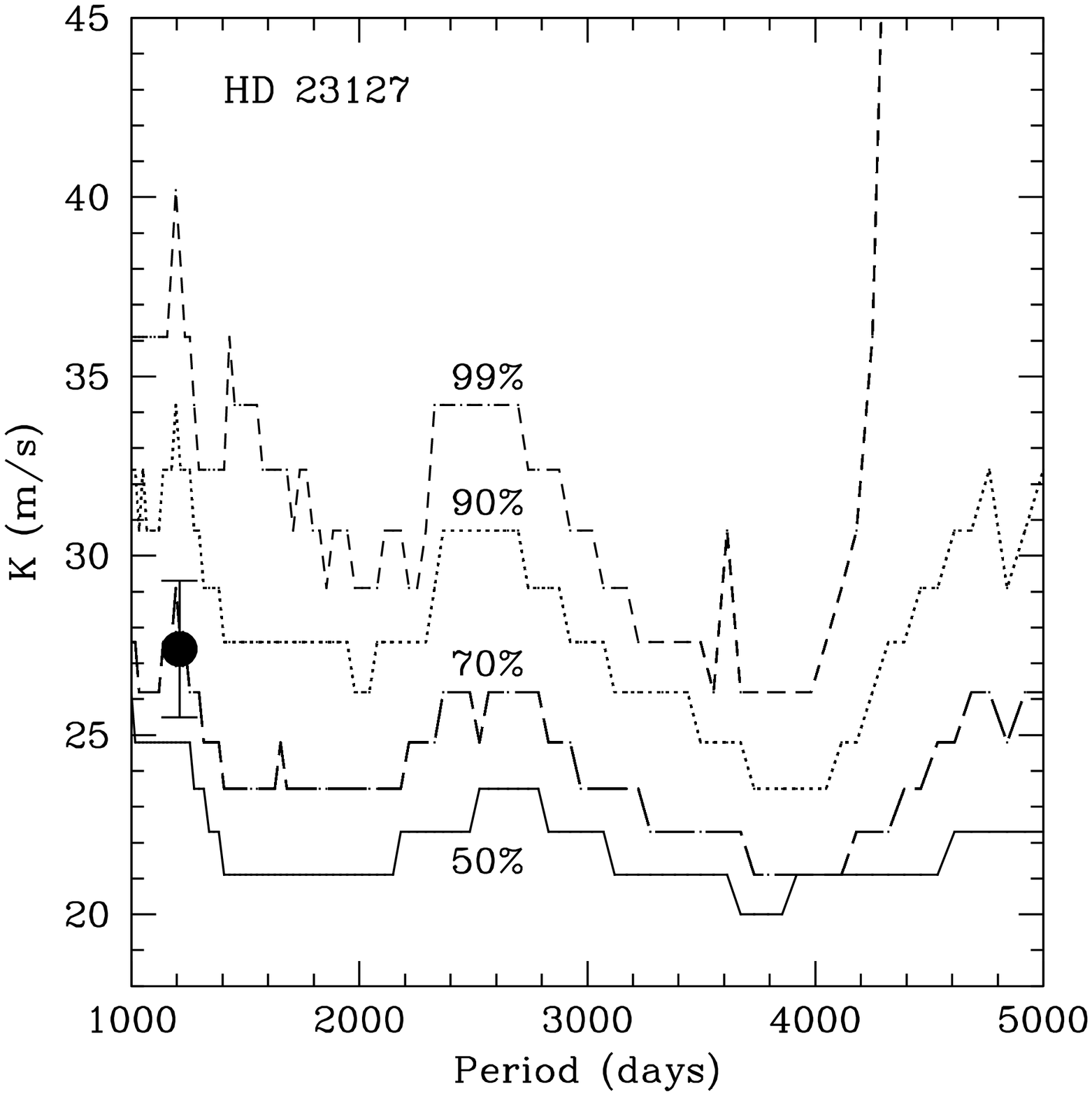}
\caption{Detection limits for GJ~832 (left) and HD~23127 (right) at 
$e=0.2$.  The known planet is plotted as a large point with error bars.  
At the specific period of each planet, the detectability is 99\% for 
GJ~832 and 70\% for HD~23127.  This shows that the automated detection 
criteria used in this work are more conservative than a human 
investigator. }
\label{getlucky}
\end{figure}

Of the 123 AAPS program targets which (1) have more than 8 years of 
data, and (2) have more than 30 observations, 25 are known to host at 
least one planet (plus one brown dwarf host: HD~164427).  The planetary 
parameters are listed in Table~\ref{planets}.  A useful sanity check is 
to ask: ``Do the detection-limit results indicate that we could have 
detected the (known) planet''?  That is, we first removed the known 
planet's orbit, then added simulated planetary signals to estimate the 
detectability of a given signal.  Applying this method to the planet 
hosts, then, serves as a simple check of the degree to which this 
detection-limit method can be trusted.  By comparing the velocity 
amplitude $K$ of the known planets in Table~\ref{planets} to the 
$\bar{K}$ detectable for those stars (Table~\ref{meankays}), we find 
that all but four of those planets pass this test at the 99\% recovery 
level: $\bar{K}<K_{planet}$.  The four exceptions are: HD~4308b, 
HD~16417b, HD~23127b, and GJ~832b.  For HD~4308 and HD~16417, this 
result is easily understood by referring to \citet{16417paper} and 
\citet{monster}.  Both of these low-amplitude, short-period planets were 
only detectable in AAT data during 48-night continuous observing blocks, 
rather than the several preceding years of more widely-spaced 
observations.  Furthermore, these short-period ($P<20$ days) planets are 
wholly different from the Jupiter analogs ($P>3000$ days) which are the 
focus of this work.  For GJ~832 and HD~23127, we show the detection 
limit expressed in terms of $K$ as a function of orbital period in 
Figure~\ref{getlucky}.  In each panel, the known planet is plotted as a 
large point with error bars.  We see that for GJ~832, the detection 
limit at the \textit{specific} period of the planet is 99\%, i.e.~the 
planet has a detectability of 99\%, though the \textit{mean} $K$ 
indicated a detectability of only $\sim$90\% (averaged over all 
periods).  Hence, these results are consistent with our robust detection 
of GJ~832b \citep{bailey09}.  For HD~23127 (right panel of 
Figure~\ref{getlucky}), the planet lies on the 70\% recovery contour; 
this example illustrates that the automated criteria are more 
conservative than a human investigator \citep{otoole07}.  This 
characteristic of our simulations also applied to the 3 planets orbiting 
61~Vir \citep{61vir}, as detailed in \citet{foreverpaper}.

\subsection{Detectability of Jupiter analogs }

\begin{figure}
\plotone{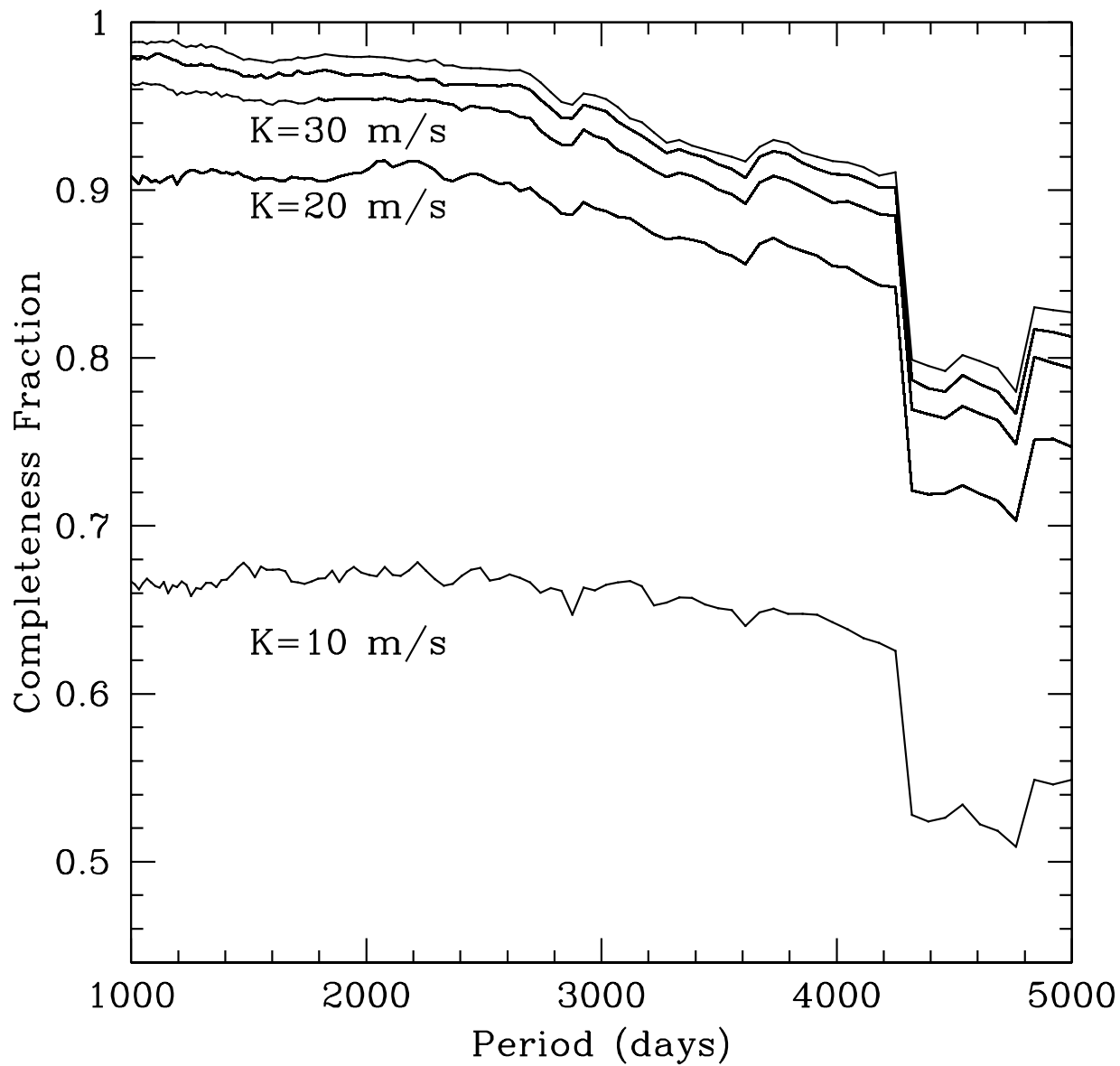}
\caption{Completeness fraction $f_{c}(P_i,K_i)$ (Equation~2) summed over 
101 AAPS stars, as a function of orbital period and radial-velocity 
amplitude $K$.  From bottom to top, the curves are for $K=$10, 20, 30, 
40, and 50 \ms.  The drop-off in completeness at $P\sim$4300 days occurs 
because that is the maximum duration of observations for any star in the 
sample. }
\label{complete1}
\end{figure}

In this work, we have adopted the definition of ``Jupiter analog'' as a 
giant planet with a long period ($P\gtsimeq$8 yr).  By this definition, 
the AAPS sample includes 3 published Jupiter analogs (HD~134987c, 
GJ~832b, and HD~160691c\footnote{In this paper, we adopt the designation 
``c'' for the outermost planet in the HD~160691 system, after 
\citet{mccarthy04}}).  So, to first order, we have a Jupiter-analog 
frequency of $3/123=2.4\%$.  However, this simple calculation assumes 
that such planets are perfectly detectable for all stars in the sample, 
which is patently false.  As we have computed circular-orbit detection 
limits at five recovery levels (99, 90, 70, 50, and 10\%), we can use 
these results to apply a rudimentary correction for the relative 
detectabilities (i.e.~the completeness) for each star.  
\citet{cumming08} detailed a similar technique to address the problem of 
incompleteness for the Keck planet search data.  Here, we adopt a simple 
approach and define the survey completeness for a given radial-velocity 
amplitude $K$ and period $P$ as:

\begin{equation}
f_{c}(P_i,K_i) = \frac{1}{N_{stars}}\sum_{j=1}^{N_{stars}} f_{R,j}(P_i,K_i),
\end{equation}

\noindent where $f_R(P,K)$ is the recovery rate as a function of $K$ at 
period $P$, and $N_{stars}$ is the total number of stars in the sample 
($N=123$).  In this way, we account for the detectabilities for each 
star individually, at each of the 100 trial periods.  We use the 
specific detection limit $K_P$ obtained for each period from the 
simulations described above, thus generating five pairs of ($K_P$, 
recovery fraction).  Then, we generate $f_R(P,K)$ for each star by 
performing a linear interpolation between the five pairs of ($K_P$, 
recovery fraction).  We can then estimate the recovery fraction 
$f_R(P,K)$ for any $P$ and $K$.  As an example, suppose we choose $K=10$ 
\ms\ and $P=1000$ days.  For HD~10700, which is extremely stable and 
well-observed, and has very tight detection limits 
($\bar{K}_{e=0.0}=3.7$\ms), a signal of 10 \ms\ would always be 
detected, and hence $f_R(1000 days, 10 \ms)$ for HD~10700 is 1.0.  For a 
star with poorer detection limits such as HD~109200 
($\bar{K}_{e=0.0}=12.3$\ms), we obtain $f_R(1000 days, 10 \ms)=0.795$.  
We can see from these examples that if all stars in the sample were 
stable and well-observed (i.e.~if selection effects, observing windows, 
and velocity jitter were unimportant), then every star would contribute 
1.0 to the sum in Equation~(2), giving a survey completeness of 1.0 
(100\%).  We could then obtain the planet frequency simply by dividing 
the number of detections by the total number of stars.  However, these 
effects are extremely important for long-term radial-velocity surveys, 
and so we use Equation~(2) to obtain a more realistic estimate of the 
completeness of our entire sample as a function of orbital period.  
Those results are shown in Figure~\ref{complete1}.  We emphasize that we 
have \textit{not} included unpublished planet candidates in our estimate 
of the frequency of Jupiter analogs.  In Figure~\ref{complete1}, we have 
summed over 101 stars, excluding 22 stars which exhibited an unusual 
artefact arising from the ``correct-period'' criterion discussed in 
Section~3.2.  Consider a data set in which an injected signal with 
$P_{in}\sim3500$ days results in $P_{out}=5000$ days being recovered by 
the periodogram.  For a recovered signal to be accepted, it must be 
within 27.7\% of the injected periodicity.  So, for $P_{in}=3612$d, 
$P_{out}=5000$d is rejected, but at $P_{in}=3671$d, $P_{out}=5000$d is 
accepted.  This resulted in signals which were ``undetectable'' even at 
large $K$ becoming eminently detectable once $P_{in}$ was within 27.7\% 
of 5000d.

The recovery fraction $f_R(P,K)$ in Equation~(2) can be used to derive a 
completeness correction for the published detections of Jupiter analogs 
in the AAPS sample.  For each of the three stars which hosts a Jupiter 
analog, we can compute $f_R(P,K)$ at the specific values of $P$ and $K$ 
for that known planet.  As we have computed detection limits for 
eccentricities of 0.0, 0.1, and 0.2, we used the results from the 
eccentricity closest to that of each detected planet (GJ~832b: $e=0.2$, 
HD~160691c: $e=0.0$, HD~134987c: $e=0.1$).  This gives the following 
results: GJ~832 -- 0.983; HD~160691 -- 1.000; HD~134987 -- 1.000.  Then, 
we compute the survey completeness $f_{c}(P,K)$ to account for the 
detectability of these specific planets around the 120 remaining stars 
that do not host a Jupiter analog.  Here, $f_{c}(P,K)$ is computed at 
the specific values of $P$ and $K$ for each Jupiter analog; those 
results are then: GJ~832 -- 0.685; HD~160691 -- 0.863; HD~134987 -- 
0.440.  The frequency of Jupiter analogs based on this sample, corrected 
for completeness (detectability), is then given by

\begin{equation}
f_{Jup} = \frac{1}{N_{stars}}\sum_{i=1}^{N_{hosts}} 
\frac{1}{f_{R,i}(P_i,K_i)f_{c}(P_i,K_i)} = 3.3\%.
\end{equation}

\noindent Here, $N_{stars}=123$ total stars in the sample, $N_{hosts}=3$ 
which host a Jupiter analog, and $f_R(P_i,K_i)$ refers to the recovery 
fractions listed above.  In addition, $f_c(P_i,K_i)$ is summed over the 
120 stars which did not host a Jupiter analog, to account for how 
detectable the three found planets would have been around the remaining 
stars in the sample.

We can also use the non-detections to compute an upper bound on the 
frequency of Jupiter analogs in the AAPS sample.  Using the recovery 
fraction $f_R(P,K)$ determined as above for each star at every trial 
period, we compute the mean of $f_R(P,K)$ over the period range 
3000-5000 days.  Thus, each of the 120 stars which does \textit{not} 
host a Jupiter analog has a mean recovery fraction $\bar{f_R}$, with an 
uncertainty equal to the standard deviation in $f_R(P,K)$ about that 
mean.  An upper bound on the frequency of Jupiter analogs can then be 
given by

\begin{equation}
\mathrm{Upper~bound}= \frac{1}{N_{nonhosts}}\sum_{i=1}^N 1-\bar{f_R},
\end{equation}

\noindent where $N_{nonhosts}$ is the number of stars without a Jupiter 
analog.  The result is an upper bound on the frequency of such planets 
with $K$ greater than the selected value.  Those results are given in 
Table~\ref{upperlimits}.

Our upper bound of 37\% for planets with $K>10$ \ms\ and $e=0.0$ in the 
period range 3000-5000 days (3~AU$<a<$6~AU) is consistent with the 
core-accretion simulations of \citet{liu09}, in which 83/220 (37\%) of 
systems resulted in at least one planet matching these characteristics.  
However, our observed frequency of such planets in this sample, 3.3\%, 
is significantly smaller.  This difference arises from the simplifying 
assumptions in the \citet{liu09} simulations.  First, each simulated 
system started with a disk mass of 1 minimum-mass solar nebula.  A 
distribution of disk masses may better approximate the formation 
environments of real planetary systems; Liu et al.~are performing 
further tests with this modification (Huigen Liu, priv.~comm.).  Second, 
the simulations proceeded for $10^7$ yr; it is possible that subsequent 
dynamical evolution such as planet-planet scattering events 
\citep{rasio96} may eject planets in real systems, which would reduce 
the observed occurence rate.

\section{Summary and Conclusions}

\citet{lineweaver03} addressed the frequency of Jupiter-like planets, 
defining ``Jupiter-like'' as those planets with $M_{Saturn}<M 
\mathrm{sin}~i < 3\Mjup$ and $3<a<9$~AU.  Based on the results of eight 
different Doppler planet surveys, they estimated that 5$\pm$2\% of 
Sun-like stars host Jupiter-like planets.  Our estimate of 3.3\% is 
consistent with theirs.  We note that this work, unlike that of 
\citet{lineweaver03}, is based on detailed simulations on a star-by-star 
basis.  This approach, while computationally intensive, is critical for 
a proper characterisation of the selection effects present in 
radial-velocity data sets \citep{monster, foreverpaper}.

The comprehensive study on detectabilities from the Keck planet search 
given in \citet{cumming08} focused on planets with periods less than 
2000 days ($a\sim$3~AU for a 1 \Msun\ star).  Those authors then 
extrapolated the simulations results to $a$=20~AU ($P$=89 years).  From 
their Table~2, which gives the giant planet ($M_p>$0.3 \Mjup) occurrence 
rates for flat and power-law extrapolations, they estimated that 
2.7$\pm$0.8\% of stars host planets between 3--6~AU.  In that range, our 
measured occurrence rate is 3.3$\pm$1.4\%.  Our results are thus 
entirely consistent with those of \citet{cumming08}.

In conclusion, we have performed detailed star-by-star simulations on 
AAPS data with a time coverage of 12 years, including the effects of 
eccentricity, in order to make a robust estimate of the frequency of 
Jupiter analogs in the Solar neighbourhood.  Based on our AAPS sample, 
we calculate that no less than 3.3\% and no more than 37.2\% of stars 
host a gas giant planet in a circular orbit between 3--6~AU.  We have 
also performed a comprehensive analysis of the automated detection 
criteria employed in our simulation method.  Salient points arising from 
this work are (1) the uncertainty in the orbital period of a planet 
increaes dramatically when less than one cycle has been observed, (2) no 
planet has been published with less than 0.7 cycles of data, and (3) 
simulating a large number of planet detections reveals that for 99\% of 
planets, the fractional uncertainty in period $\sigma_{P}/P$ is 27.7\%.  
This figure is useful for simulations in which one must determine 
whether an injected signal has been accurately redetected.

%
%

\acknowledgements

We gratefully acknowledge the UK and Australian government support of 
the Anglo-Australian Telescope through their PPARC, STFC and DIISR 
funding; STFC grant PP/C000552/1, ARC Grant DP0774000 and travel support 
from the Anglo-Australian Observatory.

We would like to thank Huigen Liu and Ji-Lin Zhou for helpful details 
about their simulation results.  We thank the AAT TAC for the generous 
allocation of telescope time to the Anglo-Australian Planet Search.  
This research has made use of NASA's Astrophysics Data System (ADS), and 
the SIMBAD database, operated at CDS, Strasbourg, France.  
R.W.~gratefully acknowledges support from a UNSW Vice-Chancellor's 
Fellowship and the UNSW Early-Career Researcher Grants program.  This 
research has made use of the Exoplanet Orbit Database and the Exoplanet 
Data Explorer at exoplanets.org.  We are also grateful to the anonymous 
referee whose comments greatly improved the clarity of this paper.


\begin{deluxetable}{lrr}
\tabletypesize{\scriptsize}
\tablecolumns{3}
\tablewidth{0pt}
\tablecaption{Summary of Radial-Velocity Data }
\tablehead{
\colhead{Star} & \colhead{$N$} & \colhead{RMS}\\
\colhead{} & \colhead{} & \colhead{(\ms)}
 }
\startdata
\label{rvdata1}
HD 142 & 65 & 17.6\tablenotemark{3} \\
HD 1581 & 87 & 3.5 \\
HD 2039 & 42 & 15.7\tablenotemark{3} \\
HD 2151 & 172 & 4.2 \\
HD 3823 & 64 & 5.9 \\
HD 4308 & 101 & 4.3\tablenotemark{3} \\
HD 7570 & 36 & 6.5 \\
HD 10180 & 31 & 6.7 \\
HD 10360 & 58 & 4.6\tablenotemark{1} \\
HD 10361 & 57 & 4.3\tablenotemark{1} \\
HD 10647 & 35 & 14.1 \\
HD 10700 & 209 & 3.6 \\
HD 11112 & 30 & 8.3\tablenotemark{1} \\
HD 13445 & 46 & 21.1\tablenotemark{1,3} \\    
HD 16417 & 103 & 3.6\tablenotemark{3} \\
HD 19632 & 33 & 26.2 \\
HD 20766 & 37 & 10.9 \\
HD 20782 & 40 & 5.8\tablenotemark{3} \\
HD 20794 & 121 & 3.2 \\
HD 20807 & 81 & 4.3 \\
HD 23127 & 38 & 12.6\tablenotemark{3} \\
HD 27442 & 74 & 7.3\tablenotemark{3} \\
HD 28255A & 59 & 7.3\tablenotemark{1} \\
HD 28255B & 39 & 24.1\tablenotemark{1} \\
HD 38382 & 32 & 5.1 \\
HD 38973 & 30 & 4.7 \\
HD 39091 & 54 & 5.6\tablenotemark{3} \\
HD 43834 & 115 & 5.1 \\
HD 44594 & 30 & 5.8 \\
HD 45701 & 30 & 5.9\tablenotemark{2} \\
HD 53705 & 115 & 4.5 \\
HD 53706 & 32 & 2.9 \\
HD 55693 & 31 & 7.2 \\
HD 59468 & 33 & 5.2 \\
HD 65907a & 53 & 6.5 \\
HD 70642 & 37 & 4.8\tablenotemark{3} \\
HD 70889 & 34 & 17.6 \\
HD 73121 & 34 & 6.2 \\
HD 73524 & 76 & 5.2 \\
HD 73526 & 33 & 8.0\tablenotemark{3} \\
HD 74868 & 33 & 7.8 \\
HD 75289 & 38 & 5.9\tablenotemark{3} \\
HD 76700 & 40 & 6.4\tablenotemark{3} \\
HD 78429 & 32 & 8.9 \\
HD 84117 & 113 & 5.5 \\
HD 88742 & 30 & 12.8 \\
HD 92987 & 44 & 5.9\tablenotemark{2} \\
HD 93385 & 36 & 6.1 \\
HD 96423 & 33 & 5.7 \\
HD 101959 & 36 & 6.4 \\
HD 102117 & 53 & 4.4\tablenotemark{3} \\
HD 102365 & 137 & 3.2 \\
HD 102438 & 42 & 4.3 \\
HD 105328 & 36 & 5.9 \\
HD 106453 & 36 & 11.2 \\
HD 107692 & 35 & 11.6 \\
HD 108147 & 55 & 19.0 \\
HD 108309 & 49 & 3.5 \\
HD 109200 & 30 & 4.6 \\
HD 114613 & 188 & 5.7 \\
HD 114853 & 43 & 5.5 \\
HD 117618 & 65 & 5.2\tablenotemark{3} \\
HD 120237 & 42 & 10.5 \\
HD 122862 & 88 & 4.4 \\
HD 125072 & 64 & 5.0 \\
HD 128620 & 89 & 3.5\tablenotemark{2} \\
HD 128621 & 126 & 3.4\tablenotemark{2} \\
HD 129060 & 37 & 37.6 \\
HD 134060 & 82 & 5.5 \\
HD 134330 & 34 & 5.7 \\
HD 134331 & 48 & 5.3\tablenotemark{1} \\
HD 134606 & 46 & 5.3 \\
HD 134987 & 60 & 2.5\tablenotemark{3} \\
HD 136352 & 134 & 4.6 \\
HD 140901 & 76 & 10.3 \\
HD 144628 & 42 & 4.0 \\
HD 147722 & 57 & 16.8 \\
HD 147723 & 60 & 9.1 \\
HD 154577 & 31 & 4.4 \\
HD 155974 & 39 & 7.5 \\
HD 156274b & 88 & 6.4\tablenotemark{1} \\
HD 160691 & 155 & 2.4\tablenotemark{3} \\
HD 161612 & 37 & 3.7 \\
HD 164427 & 42 & 6.0\tablenotemark{3} \\
HD 168871 & 56 & 4.9 \\
HD 177565 & 82 & 4.0 \\
HD 179949 & 61 & 11.1\tablenotemark{3} \\
HD 181428 & 34 & 7.9 \\
HD 183877 & 32 & 5.3 \\
HD 187085 & 49 & 5.7\tablenotemark{3} \\
HD 189567 & 73 & 5.8 \\
HD 190248 & 187 & 4.1 \\
HD 191408 & 153 & 4.2\tablenotemark{1} \\
HD 192310 & 133 & 4.0 \\
HD 192865 & 37 & 11.0 \\
HD 193193 & 44 & 5.8 \\
HD 193307 & 71 & 4.2 \\
HD 194640 & 64 & 4.8 \\
HD 196050 & 49 & 7.8\tablenotemark{3} \\
HD 196378 & 33 & 6.7 \\
HD 199190 & 41 & 3.8 \\
HD 199288 & 56 & 5.3 \\
HD 199509 & 30 & 6.6\tablenotemark{1} \\
HD 202560 & 34 & 4.9 \\
HD 204385 & 34 & 6.8 \\
HD 204961 & 32 & 5.7\tablenotemark{3} \\  
HD 207129 & 100 & 4.9 \\
HD 208487 & 41 & 6.0\tablenotemark{3} \\
HD 208998 & 32 & 8.4 \\
HD 209653 & 30 & 4.4 \\
HD 210918 & 57 & 4.6 \\
HD 211317 & 38 & 4.3 \\
HD 212168 & 34 & 4.9 \\
HD 212708 & 34 & 4.1\tablenotemark{1} \\
HD 213240 & 33 & 4.5\tablenotemark{3} \\
HD 214953 & 72 & 6.8 \\
HD 216435 & 69 & 7.0\tablenotemark{3} \\
HD 216437 & 46 & 4.9\tablenotemark{3} \\
HD 217958 & 30 & 9.4\tablenotemark{1} \\
HD 217987 & 33 & 8.9 \\
HD 219077 & 57 & 5.4\tablenotemark{2} \\
HD 221420 & 66 & 4.8\tablenotemark{1} \\
HD 223171 & 54 & 6.3 \\
\enddata
\tablenotetext{1}{Residuals to a linear fit.}
\tablenotetext{2}{Residuals to a quadratic fit.}
\tablenotetext{3}{Residuals after removal of known planet(s) orbit.}
\end{deluxetable}

\begin{deluxetable}{lcr@{$\pm$}lr@{$\pm$}lr@{$\pm$}lr@{$\pm$}lr@{$\pm$}lr@{$\pm$}
lr@{$\pm$}l}
\tabletypesize{\scriptsize}
\tablecolumns{5}
\tablewidth{0pt}
\tablecaption{Summary of Detection Limits }
\tablehead{
\colhead{Star} & \colhead{Recovery Rate} &
\multicolumn{5}{c}{Mean $K$ Detectable (\ms)} \\
\colhead{} & \colhead{(percent)} & \multicolumn{2}{c}{$e=0.0$} & 
\multicolumn{2}{c}{$e=0.1$} & \multicolumn{2}{c}{$e=0.2$}
 }
\startdata
\label{meankays}
HD 142    & 99 & 24.8 & 2.4 & 25.9 & 2.8 & 27.8 & 4.6 \\
          & 90 & 23.0 & 1.5 & 24.1 & 2.1 & 25.1 & 2.1 \\
          & 70 & 18.8 & 2.4 & 20.5 & 2.5 & 21.7 & 10.3 \\
          & 50 & 15.0 & 2.2 & 16.4 & 1.5 & 16.7 & 2.5 \\
HD 1581   & 99 & 7.2 & 9.0 & 6.4 & 5.4 & 6.0 & 3.3 \\
          & 90 & 4.5 & 1.4 & 4.9 & 2.1 & 5.7 & 3.6 \\
          & 70 & 3.1 & 0.6 & 3.5 & 0.5 & 3.6 & 0.8 \\
          & 50 & 2.6 & 1.1 & 2.8 & 1.1 & 3.0 & 1.4 \\
HD 2039   & 99 & 42.6 & 11.5 & 45.9 & 13.1 & 51.0 & 13.7 \\
          & 90 & 39.2 & 11.0 & 39.5 & 9.7 & 43.8 & 15.2 \\
          & 70 & 29.9 & 13.7 & 34.6 & 24.4 & 31.0 & 9.7 \\
          & 50 & 25.0 & 9.1 & 24.4 & 7.4 & 27.0 & 9.6 \\
HD 2151   & 99 & 6.9 & 2.9 & 7.0 & 1.6 & 8.0 & 2.6 \\
          & 90 & 6.0 & 1.2 & 6.2 & 1.1 & 6.4 & 0.8 \\
          & 70 & 5.2 & 0.8 & 5.3 & 0.6 & 5.5 & 0.6 \\
          & 50 & 4.9 & 1.4 & 4.7 & 0.7 & 4.8 & 0.7 \\
HD 3823   & 99 & 10.4 & 3.3 & 12.2 & 6.5 & 12.8 & 4.9 \\
          & 90 & 9.8 & 4.6 & 10.2 & 3.4 & 10.9 & 4.7 \\
          & 70 & 6.5 & 1.8 & 7.3 & 0.9 & 7.7 & 0.9 \\
          & 50 & 4.8 & 2.2 & 5.0 & 2.2 & 5.2 & 2.3 \\
HD 4308   & 99 & 9.1 & 8.3 & 9.4 & 7.7 & 10.8 & 8.9 \\
          & 90 & 9.4 & 9.3 & 8.0 & 5.2 & 9.0 & 6.8 \\
          & 70 & 7.0 & 3.8 & 7.3 & 4.2 & 7.1 & 3.7 \\
          & 50 & 4.9 & 2.1 & 5.2 & 2.1 & 5.2 & 1.9 \\
HD 7570   & 99 & 13.3 & 1.2 & 14.2 & 1.9 & 15.6 & 3.5 \\
          & 90 & 11.6 & 0.8 & 12.6 & 1.1 & 13.1 & 1.2 \\
          & 70 & 9.7 & 1.0 & 10.4 & 0.5 & 10.9 & 0.7 \\
          & 50 & 8.3 & 0.9 & 8.7 & 0.7 & 9.1 & 0.8 \\
HD 10180  & 99 & 16.3 & 3.1 & 17.2 & 2.3 & 19.8 & 3.4 \\
          & 90 & 14.7 & 2.3 & 15.5 & 2.7 & 17.2 & 3.6 \\ 
          & 70 & 12.1 & 1.7 & 12.8 & 1.7 & 14.0 & 2.7 \\
          & 50 & 9.9 & 2.0 & 11.0 & 1.6 & 11.3 & 2.2 \\
HD 10360  & 99 & 9.5 & 5.0 & 9.4 & 3.9 & 10.0 & 3.8 \\
          & 90 & 7.7 & 1.0 & 8.3 & 2.3 & 9.4 & 6.5 \\
          & 70 & 5.9 & 0.6 & 6.3 & 0.8 & 6.5 & 0.8 \\
          & 50 & 4.5 & 0.7 & 4.8 & 0.6 & 5.0 & 0.7 \\
HD 10361  & 99 & 7.2 & 1.2 & 7.5 & 0.8 & 7.8 & 0.9 \\
          & 90 & 6.6 & 0.5 & 6.8 & 0.6 & 7.1 & 0.7 \\
          & 70 & 5.6 & 0.5 & 5.8 & 0.5 & 6.0 & 0.5 \\
          & 50 & 4.9 & 0.5 & 5.8 & 0.5 & 5.2 & 0.4 \\
HD 10647  & 99 & 37.3 & 7.9 & 41.6 & 10.9 & 43.3 & 11.2 \\
          & 90 & 36.0 & 9.0 & 37.6 & 9.9 & 40.5 & 14.7 \\
          & 70 & 28.3 & 4.2 & 29.9 & 6.1 & 32.5 & 7.2 \\
          & 50 & 23.6 & 4.5 & 24.5 & 4.8 & 26.0 & 5.2 \\
HD 10700  & 99 & 3.7 & 0.4 & 4.6 & 2.9 & 4.2 & 0.5 \\
          & 90 & 3.4 & 0.4 & 3.6 & 0.4 & 3.8 & 0.5 \\
          & 70 & 2.1 & 0.8 & 2.1 & 0.8 & 2.3 & 0.8 \\
          & 50 & 1.6 & 0.6 & 1.6 & 0.5 & 1.6 & 0.5 \\
HD 11112  & 99 & 22.7 & 2.6 & 26.7 & 9.1 & 28.4 & 2.9 \\
          & 90 & 20.4 & 1.9 & 22.1 & 2.5 & 25.4 & 4.6 \\
          & 70 & 17.5 & 2.0 & 18.6 & 1.5 & 20.0 & 2.3 \\
          & 50 & 14.8 & 2.4 & 15.7 & 2.1 & 16.8 & 2.3 \\
HD 13445  & 99 & 41.4 & 12.4 & 44.2 & 13.4 & 48.2 & 14.0 \\
          & 90 & 37.2 & 7.6 & 37.8 & 6.1 & 41.8 & 8.1 \\
          & 70 & 33.1 & 6.8 & 32.1 & 4.2 & 36.5 & 7.5 \\
          & 50 & 24.1 & 5.1 & 22.5 & 6.1 & 23.7 & 6.3 \\
HD 16417  & 99 & 5.0 & 1.8 & 5.5 & 2.2 & 6.1 & 2.1 \\
          & 90 & 4.7 & 1.6 & 4.8 & 1.4 & 5.0 & 1.4 \\
          & 70 & 3.7 & 0.9 & 3.8 & 0.9 & 3.9 & 0.8 \\
          & 50 & 2.4 & 0.9 & 2.8 & 1.0 & 2.8 & 0.9 \\
HD 19632  & 99 & 124.1 & 36.8 & 126.2 & 36.2 & 129.8 & 36.3 \\
          & 90 & 108.8 & 28.3 & 111.7 & 33.2 & 123.3 & 48.4 \\
          & 70 & 86.5 & 27.5 & 88.5 & 30.6 & 88.7 & 23.3 \\
          & 50 & 73.0 & 22.4 & 87.0 & 35.4 & 86.9 & 29.7 \\
HD 20766  & 99 & 23.4 & 2.1 & 24.9 & 2.4 & 27.6 & 3.0 \\
          & 90 & 21.2 & 2.4 & 22.4 & 1.9 & 24.0 & 2.2 \\
          & 70 & 12.3 & 8.5 & 12.7 & 8.7 & 13.4 & 9.3 \\
          & 50 & 9.0 & 6.8 & 9.5 & 7.1 & 9.9 & 7.6 \\
HD 20782  & 99 & 11.0 & 1.0 & 12.0 & 2.3 & 12.7 & 1.1 \\
          & 90 & 10.1 & 0.7 & 10.6 & 0.9 & 11.2 & 1.2 \\
          & 70 & 8.7 & 0.6 & 9.1 & 0.6 & 9.5 & 0.6 \\
          & 50 & 7.2 & 1.0 & 7.8 & 0.4 & 8.3 & 0.6 \\
HD 20794  & 99 & 3.9 & 1.6 & 4.2 & 2.2 & 4.0 & 1.5 \\
          & 90 & 3.3 & 0.7 & 3.6 & 0.9 & 3.9 & 1.4 \\
          & 70 & 2.6 & 0.3 & 2.7 & 0.3 & 2.8 & 0.4 \\
          & 50 & 2.3 & 0.5 & 2.3 & 0.4 & 2.6 & 1.0 \\
HD 20807  & 99 & 8.9 & 5.0 & 8.7 & 3.2 & 9.2 & 2.9 \\
          & 90 & 7.0 & 2.0 & 7.6 & 2.4 & 7.8 & 2.2 \\
          & 70 & 4.8 & 1.1 & 4.9 & 0.9 & 5.2 & 1.1 \\
          & 50 & 3.3 & 1.1 & 3.3 & 1.1 & 3.3 & 1.2 \\
HD 23127  & 99 & 27.4 & 2.4 & 29.4 & 3.8 & 32.2 & 3.9 \\
          & 90 & 25.5 & 2.0 & 26.3 & 2.1 & 28.4 & 2.5 \\
          & 70 & 22.8 & 1.8 & 23.3 & 1.7 & 24.4 & 1.8 \\
          & 50 & 20.7 & 1.5 & 21.2 & 1.4 & 22.2 & 1.4 \\
HD 27442  & 99 & 10.0 & 3.8 & 9.9 & 1.2 & 11.1 & 1.7 \\
          & 90 & 8.7 & 1.0 & 9.1 & 1.2 & 9.6 & 1.1 \\
          & 70 & 7.4 & 0.9 & 7.8 & 0.7 & 8.1 & 1.1 \\
          & 50 & 6.2 & 0.9 & 6.4 & 0.8 & 6.7 & 0.8 \\
HD 28255A & 99 & 17.5 & 7.1 & 18.3 & 6.8 & 21.0 & 10.6 \\
          & 90 & 14.0 & 3.0 & 15.0 & 4.2 & 36.9 & 21.0 \\
          & 70 & 11.4 & 3.5 & 11.7 & 2.0 & 11.8 & 2.0 \\
          & 50 & 8.9 & 2.6 & 9.4 & 2.3 & 9.8 & 2.7 \\
HD 28255B & 99 & 56.4 & 11.9 & 59.5 & 12.8 & 65.8 & 14.7 \\
          & 90 & 54.7 & 13.1 & 53.5 & 10.5 & 58.3 & 12.0 \\
          & 70 & 47.6 & 7.6 & 50.9 & 11.4 & 53.8 & 16.7 \\
          & 50 & 42.1 & 7.6 & 47.1 & 5.8 & 47.0 & 8.0 \\
HD 38382  & 99 & 10.4 & 3.3 & 12.2 & 6.5 & 12.8 & 4.9 \\
          & 90 & 9.8 & 4.6 & 10.2 & 3.4 & 10.9 & 4.7 \\
          & 70 & 6.5 & 1.8 & 7.3 & 0.9 & 7.7 & 0.9 \\
          & 50 & 4.8 & 2.2 & 5.0 & 2.2 & 5.2 & 2.3 \\
HD 38973  & 99 & 13.8 & 4.1 & 15.4 & 2.2 & 18.2 & 3.6 \\
          & 90 & 11.6 & 1.1 & 12.6 & 1.6 & 14.0 & 1.7 \\
          & 70 & 9.8 & 0.6 & 10.3 & 1.0 & 11.0 & 1.1 \\
          & 50 & 8.2 & 0.6 & 8.7 & 0.6 & 9.1 & 0.6 \\
HD 39091  & 99 & 9.7 & 3.7 & 10.3 & 3.5 & 10.4 & 2.1 \\
          & 90 & 8.3 & 1.3 & 8.7 & 1.6 & 9.3 & 1.8 \\
          & 70 & 6.7 & 0.9 & 7.3 & 1.1 & 7.3 & 1.0 \\
          & 50 & 4.9 & 1.5 & 5.6 & 1.1 & 5.9 & 1.2 \\
HD 43834  & 99 & 8.8 & 4.5 & 9.3 & 6.0 & 9.9 & 6.6 \\
          & 90 & 7.5 & 1.8 & 7.9 & 2.3 & 8.5 & 3.1 \\
          & 70 & 6.2 & 1.5 & 6.6 & 1.6 & 6.9 & 1.7 \\
          & 50 & 4.2 & 1.2 & 4.5 & 1.0 & 4.8 & 1.0 \\
HD 44594  & 99 & 16.2 & 5.5 & 16.3 & 1.9 & 18.3 & 2.1 \\
          & 90 & 13.7 & 2.2 & 14.8 & 2.6 & 15.4 & 1.5 \\
          & 70 & 11.7 & 1.0 & 12.8 & 1.6 & 13.7 & 3.0 \\
          & 50 & 10.4 & 0.6 & 10.9 & 0.8 & 11.6 & 1.0 \\
HD 45701  & 99 & 22.7 & 9.9 & 24.6 & 7.1 & 26.2 & 7.5 \\
          & 90 & 19.5 & 6.1 & 22.6 & 11.1 & 21.1 & 7.0 \\
          & 70 & 15.3 & 4.2 & 16.3 & 4.1 & 17.0 & 4.4 \\
          & 50 & 12.5 & 1.9 & 13.0 & 1.9 & 13.8 & 2.1 \\
HD 53705  & 99 & 6.0 & 1.4 & 6.3 & 1.2 & 6.8 & 1.1 \\
          & 90 & 5.9 & 1.7 & 6.1 & 2.3 & 6.4 & 1.9 \\
          & 70 & 5.3 & 1.0 & 5.5 & 1.1 & 5.9 & 1.4 \\
          & 50 & 4.4 & 0.5 & 4.6 & 0.6 & 4.8 & 0.6 \\
HD 53706  & 99 & 6.0 & 0.5 & 6.7 & 0.7 & 7.7 & 1.5 \\
          & 90 & 5.5 & 0.2 & 5.7 & 0.3 & 6.2 & 0.4 \\
          & 70 & 4.9 & 0.3 & 5.1 & 0.3 & 5.4 & 0.3 \\
          & 50 & 4.4 & 0.4 & 4.6 & 0.3 & 4.8 & 0.3 \\
HD 55693  & 99 & 18.5 & 2.7 & 20.6 & 3.5 & 24.7 & 6.2 \\
          & 90 & 16.6 & 2.3 & 17.8 & 2.4 & 19.8 & 3.0 \\
          & 70 & 13.9 & 1.8 & 14.8 & 1.5 & 15.9 & 1.6 \\
          & 50 & 11.8 & 2.1 & 12.5 & 1.9 & 13.3 & 2.2 \\
HD 59468  & 99 & 11.4 & 1.2 & 12.1 & 1.3 & 13.4 & 1.4 \\
          & 90 & 10.1 & 0.7 & 10.6 & 0.9 & 11.4 & 0.9 \\
          & 70 & 8.4 & 0.8 & 9.1 & 0.5 & 9.7 & 0.5 \\
          & 50 & 7.0 & 1.2 & 7.4 & 1.0 & 7.8 & 1.1 \\
HD 65907a & 99 & 10.1 & 1.5 & 10.8 & 1.6 & 11.8 & 2.2 \\
          & 90 & 8.8 & 1.0 & 9.6 & 1.5 & 10.1 & 1.5 \\
          & 70 & 6.0 & 1.8 & 7.1 & 0.8 & 7.5 & 0.7 \\
          & 50 & 4.3 & 2.0 & 4.4 & 2.1 & 4.6 & 2.1 \\
HD 70642  & 99 & 10.0 & 1.0 & 11.2 & 2.0 & 12.0 & 1.7 \\
          & 90 & 8.8 & 0.7 & 9.5 & 0.9 & 10.3 & 1.0 \\
          & 70 & 7.4 & 0.6 & 7.8 & 0.5 & 8.2 & 0.7 \\
          & 50 & 6.5 & 0.6 & 6.8 & 0.4 & 7.2 & 0.3 \\
HD 70889  & 99 & 58.5 & 25.7 & 59.7 & 21.6 & 74.2 & 29.8 \\
          & 90 & 51.1 & 17.4 & 54.2 & 21.2 & 58.4 & 20.9 \\
          & 70 & 39.6 & 7.6 & 41.6 & 9.1 & 45.1 & 10.9 \\
          & 50 & 31.9 & 8.0 & 32.4 & 6.3 & 33.4 & 5.7 \\
HD 73121  & 99 & 12.3 & 1.1 & 13.3 & 0.9 & 15.1 & 2.7 \\
          & 90 & 11.1 & 0.5 & 11.7 & 0.7 & 12.7 & 0.7 \\
          & 70 & 9.7 & 1.0 & 10.0 & 0.6 & 10.7 & 0.8 \\
          & 50 & 8.6 & 1.2 & 8.9 & 1.0 & 9.4 & 1.0 \\
HD 73524  & 99 & 10.7 & 4.0 & 11.7 & 4.5 & 12.2 & 3.3 \\
          & 90 & 8.8 & 1.8 & 9.3 & 2.0 & 10.3 & 3.7 \\
          & 70 & 6.9 & 0.9 & 7.3 & 0.7 & 7.8 & 0.9 \\
          & 50 & 5.2 & 0.8 & 5.5 & 0.8 & 5.7 & 0.8 \\
HD 73526  & 99 & 22.8 & 6.2 & 25.3 & 8.3 & 29.1 & 11.0 \\
          & 90 & 19.5 & 3.4 & 21.4 & 5.1 & 23.6 & 6.7 \\
          & 70 & 17.8 & 3.1 & 18.6 & 4.1 & 19.2 & 3.0 \\
          & 50 & 15.2 & 1.4 & 15.8 & 1.8 & 16.9 & 2.6 \\
HD 74868  & 99 & 18.0 & 1.6 & 19.4 & 2.1 & 21.8 & 3.1 \\
          & 90 & 16.2 & 1.3 & 17.2 & 1.4 & 18.7 & 1.7 \\
          & 70 & 12.9 & 2.2 & 14.3 & 1.4 & 14.8 & 1.7 \\
          & 50 & 9.5 & 3.6 & 10.2 & 3.4 & 10.9 & 3.6 \\
HD 75289  & 99 & 11.4 & 1.0 & 12.3 & 1.4 & 13.8 & 1.6 \\
          & 90 & 10.5 & 0.7 & 11.1 & 0.8 & 11.9 & 1.0 \\
          & 70 & 8.7 & 1.0 & 9.6 & 1.0 & 9.8 & 1.1 \\
          & 50 & 7.3 & 1.3 & 8.0 & 1.4 & 8.4 & 1.5 \\
HD 76700  & 99 & 14.3 & 2.0 & 16.3 & 4.3 & 17.6 & 3.7 \\
          & 90 & 13.4 & 1.8 & 13.9 & 2.0 & 15.1 & 2.4 \\
          & 70 & 12.2 & 4.4 & 11.4 & 2.1 & 12.4 & 2.3 \\
          & 50 & 9.9 & 3.2 & 10.2 & 3.4 & 10.6 & 3.6 \\
HD 78429  & 99 & 13.3 & 1.2 & 14.2 & 1.9 & 15.6 & 3.5 \\
          & 90 & 11.6 & 0.8 & 12.6 & 1.1 & 13.1 & 1.2 \\
          & 70 & 9.7 & 1.0 & 10.4 & 0.5 & 10.9 & 0.7 \\
          & 50 & 8.3 & 0.9 & 8.7 & 0.7 & 9.1 & 0.8 \\
HD 84117  & 99 & 8.4 & 2.9 & 8.9 & 3.5 & 9.1 & 3.3 \\
          & 90 & 7.4 & 1.2 & 7.8 & 1.4 & 8.3 & 1.9 \\
          & 70 & 6.4 & 1.0 & 6.8 & 1.0 & 7.2 & 1.1 \\
          & 50 & 5.8 & 0.8 & 6.0 & 0.8 & 6.3 & 0.9 \\
HD 88742  & 99 & 31.0 & 1.5 & 33.8 & 2.6 & 40.5 & 9.1 \\
          & 90 & 28.5 & 1.1 & 29.9 & 1.5 & 32.9 & 2.6 \\
          & 70 & 23.5 & 3.0 & 24.5 & 2.5 & 26.4 & 2.9 \\
          & 50 & 16.2 & 5.7 & 17.7 & 5.1 & 18.8 & 5.5 \\
HD 92987  & 99 & 9.6 & 0.6 & 10.2 & 1.0 & 10.5 & 0.7 \\
          & 90 & 8.9 & 0.5 & 9.1 & 0.4 & 9.7 & 0.8 \\
          & 70 & 8.1 & 0.4 & 8.2 & 0.3 & 8.8 & 0.3 \\
          & 50 & 7.2 & 0.3 & 7.3 & 0.3 & 7.9 & 0.5 \\
HD 93385  & 99 & 13.4 & 1.3 & 14.9 & 1.8 & 17.1 & 3.0 \\
          & 90 & 11.9 & 1.1 & 12.7 & 1.1 & 13.9 & 1.5 \\
          & 70 & 9.9 & 1.0 & 10.5 & 0.6 & 11.2 & 1.1 \\
          & 50 & 8.7 & 1.5 & 9.1 & 1.3 & 9.5 & 1.3 \\
HD 96423  & 99 & 12.8 & 1.0 & 13.5 & 1.2 & 16.3 & 4.7 \\
          & 90 & 11.6 & 0.8 & 12.4 & 0.8 & 13.2 & 0.9 \\
          & 70 & 10.4 & 0.5 & 10.9 & 0.4 & 11.5 & 0.5 \\
          & 50 & 9.5 & 0.5 & 9.9 & 0.4 & 10.4 & 0.4 \\
HD 101959 & 99 & 13.2 & 1.2 & 14.3 & 1.5 & 16.0 & 1.8 \\
          & 90 & 11.9 & 0.9 & 11.9 & 0.9 & 13.7 & 1.2 \\
          & 70 & 10.5 & 0.7 & 10.5 & 0.7 & 10.9 & 0.6 \\
          & 50 & 9.1 & 0.6 & 9.5 & 0.7 & 9.9 & 0.6 \\
HD 102117 & 99 & 7.0 & 0.7 & 7.6 & 0.9 & 8.5 & 0.6 \\
          & 90 & 6.6 & 0.5 & 6.7 & 0.5 & 7.3 & 0.6 \\
          & 70 & 6.1 & 0.7 & 6.2 & 1.0 & 6.4 & 0.9 \\
          & 50 & 5.4 & 0.7 & 5.4 & 0.5 & 5.7 & 0.6 \\
HD 102365 & 99 & 3.8 & 0.7 & 4.1 & 1.0 & 4.9 & 1.9 \\
          & 90 & 3.5 & 0.7 & 3.7 & 0.7 & 4.1 & 1.2 \\
          & 70 & 3.1 & 0.6 & 3.4 & 0.7 & 3.5 & 0.7 \\
          & 50 & 2.4 & 0.4 & 2.6 & 0.2 & 2.7 & 0.3 \\
HD 102438 & 99 & 7.5 & 0.8 & 8.0 & 1.3 & 9.3 & 3.9 \\
          & 90 & 6.8 & 0.8 & 7.3 & 0.8 & 7.5 & 1.3 \\
          & 70 & 5.1 & 0.6 & 5.8 & 0.6 & 5.8 & 0.7 \\
          & 50 & 3.8 & 0.7 & 4.2 & 0.5 & 4.3 & 0.6 \\
HD 105328 & 99 & 11.5 & 0.9 & 12.3 & 1.0 & 13.6 & 2.1 \\
          & 90 & 10.5 & 0.6 & 10.9 & 0.6 & 11.8 & 0.8 \\
          & 70 & 8.9 & 0.8 & 9.5 & 0.3 & 9.8 & 0.6 \\
          & 50 & 7.5 & 1.3 & 7.9 & 1.1 & 8.2 & 1.2 \\
HD 106453 & 99 & 24.1 & 6.0 & 25.5 & 3.1 & 29.0 & 3.7 \\
          & 90 & 22.1 & 3.9 & 23.2 & 4.1 & 23.8 & 1.4 \\
          & 70 & 19.0 & 3.0 & 19.4 & 2.1 & 21.4 & 4.4 \\
          & 50 & 17.0 & 2.7 & 18.9 & 5.6 & 20.8 & 8.7 \\
HD 107692 & 99 & 29.2 & 8.3 & 29.0 & 2.4 & 32.5 & 3.7 \\
          & 90 & 24.8 & 4.2 & 26.8 & 4.6 & 29.5 & 6.7 \\
          & 70 & 17.8 & 3.8 & 19.7 & 4.1 & 20.5 & 4.6 \\
          & 50 & 14.8 & 3.2 & 15.4 & 3.1 & 16.4 & 3.3 \\
HD 108147 & 99 & 42.5 & 10.4 & 44.0 & 8.7 & 49.5 & 10.4 \\
          & 90 & 41.7 & 15.0 & 40.2 & 8.1 & 42.1 & 6.5 \\
          & 70 & 41.7 & 22.4 & 34.5 & 8.5 & 36.0 & 6.3 \\
          & 50 & 31.5 & 6.4 & 33.4 & 7.0 & 35.6 & 7.9 \\
HD 108309 & 99 & 5.4 & 0.8 & 5.6 & 1.3 & 5.7 & 0.7 \\
          & 90 & 4.9 & 0.4 & 5.2 & 0.5 & 5.9 & 3.1 \\
          & 70 & 4.1 & 0.3 & 4.3 & 0.3 & 4.3 & 0.3 \\
          & 50 & 3.5 & 0.4 & 3.7 & 0.4 & 3.8 & 0.5 \\
HD 109200 & 99 & 12.3 & 1.0 & 13.6 & 1.5 & 16.4 & 4.8 \\
          & 90 & 10.7 & 0.7 & 11.4 & 0.7 & 12.9 & 1.1 \\
          & 70 & 9.5 & 0.6 & 9.8 & 0.4 & 10.6 & 0.4 \\
          & 50 & 8.3 & 0.5 & 8.6 & 0.4 & 9.3 & 0.4 \\
HD 114613 & 99 & 7.5 & 3.7 & 7.9 & 4.1 & 8.5 & 4.3 \\
          & 90 & 6.8 & 3.3 & 7.1 & 3.5 & 7.2 & 3.4 \\
          & 70 & 5.4 & 2.4 & 5.5 & 2.5 & 5.6 & 2.5 \\
          & 50 & 4.4 & 1.9 & 4.4 & 1.8 & 4.6 & 1.9 \\
HD 114853 & 99 & 9.1 & 0.7 & 9.5 & 0.8 & 10.5 & 0.9 \\
          & 90 & 8.5 & 0.5 & 8.9 & 0.6 & 9.3 & 0.3 \\
          & 70 & 7.5 & 0.5 & 7.9 & 0.4 & 8.3 & 0.5 \\
          & 50 & 6.6 & 0.8 & 7.1 & 0.5 & 7.2 & 0.7 \\
HD 117618 & 99 & 7.8 & 1.4 & 8.5 & 2.3 & 10.0 & 5.5 \\
          & 90 & 7.4 & 1.4 & 7.5 & 0.9 & 8.1 & 1.4 \\
          & 70 & 6.1 & 1.0 & 6.3 & 1.3 & 7.1 & 1.8 \\
          & 50 & 4.7 & 1.3 & 4.9 & 1.4 & 5.4 & 1.4 \\
HD 120237 & 99 & 18.9 & 2.0 & 20.0 & 2.4 & 21.6 & 2.8 \\
          & 90 & 17.2 & 1.5 & 18.2 & 1.9 & 19.3 & 2.0 \\
          & 70 & 14.2 & 1.7 & 15.1 & 1.0 & 15.7 & 1.2 \\
          & 50 & 11.9 & 2.3 & 12.5 & 1.5 & 12.8 & 1.9 \\
HD 122862 & 99 & 8.3 & 6.4 & 7.8 & 3.8 & 8.0 & 3.0 \\
          & 90 & 6.5 & 2.5 & 6.5 & 1.9 & 6.6 & 1.7 \\
          & 70 & 5.0 & 0.8 & 5.1 & 0.7 & 5.3 & 0.5 \\
          & 50 & 4.4 & 0.7 & 4.5 & 0.5 & 4.7 & 0.5 \\
HD 125072 & 99 & 8.7 & 1.2 & 9.3 & 1.3 & 9.9 & 1.3 \\
          & 90 & 7.2 & 1.9 & 8.1 & 1.0 & 8.4 & 1.0 \\
          & 70 & 5.1 & 2.1 & 5.7 & 2.0 & 5.6 & 1.9 \\
          & 50 & 3.3 & 2.0 & 3.8 & 2.0 & 3.6 & 2.0 \\
HD 128620 & 99 & 6.4 & 2.2 & 7.5 & 3.6 & 9.2 & 7.5 \\
          & 90 & 5.2 & 1.7 & 5.4 & 1.4 & 6.0 & 2.0 \\
          & 70 & 3.6 & 1.1 & 3.9 & 1.1 & 4.0 & 1.3 \\
          & 50 & 2.3 & 1.1 & 3.1 & 1.1 & 2.7 & 1.0 \\
HD 128621 & 99 & 6.2 & 1.1 & 6.9 & 1.6 & 8.1 & 3.7 \\
          & 90 & 5.5 & 0.8 & 5.7 & 1.0 & 6.1 & 1.1 \\
          & 70 & 4.2 & 0.7 & 4.2 & 0.7 & 4.4 & 0.7 \\
          & 50 & 3.1 & 1.0 & 2.9 & 1.0 & 3.1 & 0.9 \\
HD 129060 & 99 & 98.2 & 21.5 & 104.6 & 24.3 & 116.8 & 28.4 \\
          & 90 & 97.2 & 25.9 & 93.7 & 16.2 & 101.9 & 17.3 \\
          & 70 & 79.8 & 14.6 & 81.8 & 16.8 & 87.2 & 17.6 \\
          & 50 & 75.8 & 24.1 & 76.7 & 21.5 & 85.8 & 26.5 \\
HD 134060 & 99 & 9.5 & 6.3 & 9.6 & 3.5 & 10.2 & 2.9 \\
          & 90 & 7.6 & 2.6 & 7.9 & 1.7 & 8.2 & 1.3 \\
          & 70 & 5.6 & 0.8 & 6.0 & 0.7 & 6.2 & 0.7 \\
          & 50 & 4.4 & 1.1 & 4.8 & 1.0 & 4.7 & 1.1 \\
HD 134330 & 99 & 12.9 & 1.3 & 13.9 & 1.8 & 15.2 & 1.7 \\
          & 90 & 11.6 & 1.1 & 12.3 & 1.3 & 13.3 & 1.7 \\
          & 70 & 10.1 & 0.8 & 10.5 & 0.7 & 11.0 & 0.9 \\
          & 50 & 8.3 & 0.7 & 8.6 & 0.8 & 9.4 & 0.7 \\
HD 134331 & 99 & 9.4 & 4.5 & 9.2 & 0.5 & 9.8 & 0.7 \\
          & 90 & 8.6 & 2.3 & 9.0 & 2.4 & 8.8 & 0.4 \\
          & 70 & 7.2 & 0.5 & 7.6 & 0.9 & 7.9 & 1.0 \\
          & 50 & 6.0 & 0.8 & 6.5 & 0.8 & 6.6 & 0.7 \\
HD 134606 & 99 & 8.8 & 1.8 & 9.3 & 2.3 & 9.7 & 0.8 \\
          & 90 & 7.7 & 0.7 & 8.1 & 1.0 & 8.4 & 0.5 \\
          & 70 & 6.6 & 0.7 & 7.1 & 0.6 & 7.5 & 0.8 \\
          & 50 & 5.8 & 0.6 & 6.1 & 0.6 & 6.3 & 0.6 \\
HD 134987 & 99 & 3.8 & 0.4 & 4.1 & 0.5 & 4.4 & 0.7 \\
          & 90 & 3.4 & 0.4 & 3.6 & 0.4 & 3.8 & 0.4 \\
          & 70 & 2.8 & 0.5 & 3.0 & 0.5 & 3.1 & 0.6 \\
          & 50 & 2.3 & 0.7 & 2.4 & 0.8 & 2.6 & 0.8 \\
HD 136352 & 99 & 7.0 & 3.3 & 6.9 & 1.1 & 7.7 & 1.9 \\
          & 90 & 6.1 & 1.4 & 7.0 & 3.6 & 6.5 & 0.8 \\
          & 70 & 5.0 & 0.6 & 5.4 & 0.7 & 5.7 & 0.8 \\
          & 50 & 4.3 & 0.3 & 4.7 & 0.5 & 4.8 & 0.5 \\
HD 140901 & 99 & 19.0 & 8.2 & 18.5 & 4.6 & 20.6 & 7.1 \\
          & 90 & 17.2 & 4.5 & 18.3 & 7.4 & 18.1 & 5.2 \\
          & 70 & 13.3 & 2.3 & 13.5 & 2.4 & 14.1 & 2.6 \\
          & 50 & 9.8 & 2.8 & 9.8 & 3.6 & 10.4 & 3.1 \\
HD 144628 & 99 & 7.1 & 1.3 & 7.3 & 0.7 & 8.1 & 1.1 \\
          & 90 & 6.5 & 0.4 & 6.7 & 0.4 & 7.1 & 0.5 \\
          & 70 & 5.6 & 0.4 & 6.0 & 0.6 & 6.1 & 0.5 \\
          & 50 & 4.7 & 0.4 & 5.0 & 0.3 & 5.2 & 0.3 \\
HD 147722 & 99 & 25.4 & 2.2 & 26.5 & 2.8 & 28.8 & 3.9 \\
          & 90 & 23.9 & 1.8 & 24.5 & 2.2 & 25.3 & 2.5 \\
          & 70 & 22.9 & 10.0 & 21.1 & 1.0 & 21.7 & 1.3 \\
          & 50 & 19.2 & 2.0 & 19.6 & 1.9 & 20.5 & 2.1 \\
HD 147723 & 99 & 15.3 & 2.1 & 15.9 & 2.4 & 17.1 & 2.6 \\
          & 90 & 14.2 & 1.9 & 14.8 & 2.1 & 15.5 & 2.5 \\
          & 70 & 11.9 & 1.2 & 12.4 & 1.2 & 13.0 & 1.6 \\
          & 50 & 9.9  & 1.6 & 10.5 & 1.4 & 11.1 & 1.6 \\
HD 154577 & 99 & 9.9 & 0.7 & 10.8 & 1.3 & 12.7 & 3.7 \\
          & 90 & 9.0 & 0.6 & 9.6 & 0.8 & 10.5 & 1.1 \\
          & 70 & 7.9 & 0.7 & 8.4 & 0.6 & 9.0 & 0.8 \\
          & 50 & 7.1 & 0.7 & 7.4 & 0.6 & 7.9 & 0.8 \\
HD 155974 & 99 & 14.4 & 1.0 & 15.3 & 1.2 & 16.8 & 1.7 \\
          & 90 & 13.4 & 0.8 & 13.9 & 0.8 & 14.9 & 1.1 \\
          & 70 & 12.1 & 1.3 & 12.5 & 1.2 & 13.8 & 3.5 \\
          & 50 & 10.8 & 1.2 & 11.1 & 0.9 & 11.9 & 1.8 \\
HD 156274b & 99 & 12.2 & 9.4 & 13.6 & 10.3 & 16.9 & 17.0 \\
          & 90 & 8.3 & 2.7 & 9.4 & 3.7 & 10.9 & 6.3 \\
          & 70 & 6.3 & 1.5 & 7.1 & 1.0 & 7.2 & 1.3 \\
          & 50 & 5.3 & 1.7 & 5.5 & 1.5 & 5.7 & 1.6 \\
HD 160691 & 99 & 2.3 & 0.3 & 2.6 & 0.9 & 2.8 & 1.1 \\
          & 90 & 2.0 & 0.2 & 2.2 & 0.2 & 2.3 & 0.4 \\
          & 70 & 1.8 & 0.4 & 1.8 & 0.2 & 1.9 & 0.2 \\
          & 50 & 1.5 & 0.2 & 1.6 & 0.2 & 1.6 & 0.4 \\
HD 161612 & 99 & 7.1 & 0.5 & 7.5 & 0.7 & 8.6 & 0.9 \\
          & 90 & 6.7 & 0.7 & 6.9 & 0.4 & 7.2 & 0.4 \\
          & 70 & 5.8 & 0.7 & 6.4 & 0.7 & 6.5 & 1.2 \\
          & 50 & 5.3 & 0.5 & 5.5 & 0.5 & 5.7 & 0.6 \\
HD 164427 & 99 & 15.9 & 8.5 & 18.0 & 17.5 & 13.9 & 3.1 \\
          & 90 & 12.6 & 3.6 & 13.9 & 4.5 & 16.0 & 6.2 \\
          & 70 & 11.5 & 6.4 & 13.0 & 6.7 & 13.5 & 8.0 \\
          & 50 & 13.4 & 10.4 & 11.9 & 5.0 & 12.8 & 6.8 \\
HD 168871 & 99 & 10.4 & 8.4 & 10.8 & 7.5 & 11.1 & 7.1 \\
          & 90 & 8.9 & 4.4 & 9.2 & 4.9 & 8.9 & 3.4 \\
          & 70 & 5.4 & 2.1 & 6.4 & 1.7 & 6.5 & 3.1 \\
          & 50 & 3.3 & 1.6 & 3.7 & 1.7 & 3.9 & 1.9 \\
HD 177565 & 99 & 7.9 & 5.0 & 7.1 & 1.5 & 7.6 & 1.4 \\
          & 90 & 6.6 & 2.8 & 6.5 & 1.2 & 6.9 & 1.6 \\
          & 70 & 3.6 & 1.8 & 3.7 & 1.9 & 4.1 & 1.8 \\
          & 50 & 2.3 & 1.5 & 2.4 & 1.5 & 2.5 & 1.5 \\
HD 179949 & 99 & 21.8 & 11.1 & 20.9 & 6.4 & 22.1 & 4.6 \\
          & 90 & 19.1 & 5.6 & 20.8 & 9.3 & 19.5 & 2.6 \\
          & 70 & 15.3 & 2.6 & 15.4 & 2.8 & 16.9 & 2.6 \\
          & 50 & 11.1 & 2.4 & 9.9 & 2.2 & 12.2 & 2.5 \\
HD 181428 & 99 & 18.4 & 1.8 & 19.8 & 2.1 & 22.1 & 3.1 \\
          & 90 & 16.8 & 1.3 & 17.4 & 1.2 & 18.6 & 1.3 \\
          & 70 & 14.2 & 1.1 & 15.5 & 1.8 & 15.9 & 2.5 \\
          & 50 & 12.5 & 1.1 & 12.9 & 1.0 & 13.8 & 1.3 \\
HD 183877 & 99 & 12.5 & 1.7 & 14.0 & 2.6 & 16.2 & 4.0 \\
          & 90 & 11.6 & 1.9 & 12.1 & 2.4 & 13.6 & 5.9 \\
          & 70 & 9.9 & 1.0 & 10.6 & 1.4 & 11.4 & 2.3 \\
          & 50 & 8.6 & 1.0 & 9.1 & 1.1 & 9.7 & 1.4 \\
HD 187085 & 99 & 8.9 & 0.6 & 9.2 & 0.6 & 9.9 & 1.0 \\
          & 90 & 8.4 & 0.5 & 8.7 & 0.5 & 9.0 & 0.5 \\
          & 70 & 7.5 & 0.9 & 7.5 & 0.2 & 7.8 & 0.4 \\
          & 50 & 6.8 & 0.5 & 7.1 & 0.7 & 7.3 & 0.4 \\
HD 189567 & 99 & 10.5 & 4.2 & 10.4 & 3.2 & 11.2 & 3.1 \\
          & 90 & 8.5 & 2.0 & 9.0 & 2.2 & 9.5 & 3.0 \\
          & 70 & 7.0 & 1.3 & 7.6 & 1.2 & 7.8 & 1.3 \\
          & 50 & 5.9 & 1.9 & 6.4 & 1.3 & 6.6 & 1.6 \\
HD 190248 & 99 & 4.9 & 1.4 & 5.6 & 2.7 & 5.8 & 2.1 \\
          & 90 & 4.3 & 0.9 & 4.5 & 0.9 & 4.8 & 1.2 \\
          & 70 & 3.5 & 0.5 & 3.8 & 0.7 & 4.0 & 0.6 \\
          & 50 & 2.9 & 0.6 & 3.2 & 0.5 & 3.2 & 0.4 \\
HD 191408 & 99 & 4.4 & 0.7 & 4.7 & 0.8 & 5.1 & 0.8 \\
          & 90 & 4.1 & 0.6 & 4.2 & 0.6 & 4.5 & 0.6 \\
          & 70 & 3.4 & 0.3 & 3.5 & 0.2 & 3.7 & 0.3 \\
          & 50 & 2.8 & 0.3 & 2.9 & 0.2 & 3.0 & 0.3 \\
HD 192310 & 99 & 3.4 & 1.8 & 3.6 & 1.9 & 3.8 & 2.0 \\
          & 90 & 3.1 & 1.6 & 3.2 & 1.7 & 3.4 & 1.7 \\
          & 70 & 2.6 & 1.3 & 2.7 & 1.3 & 2.8 & 1.4 \\
          & 50 & 2.4 & 1.2 & 2.4 & 1.3 & 2.5 & 1.3 \\
HD 192865 & 99 & 28.4 & 6.9 & 31.2 & 8.6 & 35.9 & 11.0 \\
          & 90 & 26.3 & 6.2 & 27.8 & 6.6 & 32.5 & 11.4 \\
          & 70 & 21.8 & 3.4 & 23.0 & 3.1 & 24.4 & 3.6 \\
          & 50 & 17.9 & 3.0 & 19.5 & 2.7 & 19.7 & 3.4 \\
HD 193193 & 99 & 17.1 & 14.5 & 18.4 & 15.5 & 19.5 & 17.5 \\
          & 90 & 15.0 & 10.8 & 16.7 & 14.5 & 17.4 & 14.8 \\
          & 70 & 13.1 & 12.3 & 12.8 & 11.5 & 11.0 & 4.3 \\
          & 50 & 9.0 & 3.2 & 8.9 & 2.0 & 9.9 & 3.6 \\
HD 193307 & 99 & 9.4 & 7.6 & 9.2 & 4.0 & 11.0 & 6.0 \\
          & 90 & 7.0 & 2.6 & 6.8 & 1.6 & 6.8 & 1.4 \\
          & 70 & 5.1 & 0.8 & 5.2 & 0.8 & 5.4 & 0.9 \\
          & 50 & 4.2 & 1.0 & 4.4 & 0.8 & 4.4 & 0.9 \\
HD 194640 & 99 & 10.4 & 9.1 & 8.9 & 3.3 & 8.7 & 1.3 \\
          & 90 & 7.4 & 1.1 & 7.5 & 0.7 & 7.9 & 0.8 \\
          & 70 & 6.5 & 0.7 & 6.6 & 0.8 & 6.9 & 0.9 \\
          & 50 & 6.2 & 1.0 & 6.3 & 0.9 & 6.5 & 1.0 \\
HD 196050 & 99 & 13.6 & 1.7 & 14.3 & 2.0 & 15.7 & 1.9 \\
          & 90 & 12.9 & 1.4 & 13.4 & 1.7 & 13.9 & 1.6 \\
          & 70 & 11.6 & 1.8 & 12.1 & 1.2 & 12.2 & 1.1 \\
          & 50 & 9.5 & 1.3 & 10.9 & 1.5 & 11.2 & 3.0 \\
HD 196378 & 99 & 15.3 & 1.1 & 16.8 & 1.6 & 18.7 & 1.1 \\
          & 90 & 14.0 & 0.9 & 15.0 & 1.1 & 16.4 & 1.7 \\
          & 70 & 12.7 & 1.9 & 13.2 & 0.8 & 14.1 & 1.1 \\
          & 50 & 12.0 & 1.7 & 12.3 & 1.6 & 13.1 & 1.9 \\
HD 199190 & 99 & 6.8 & 0.6 & 7.2 & 1.6 & 7.7 & 0.5 \\
          & 90 & 6.4 & 0.5 & 6.6 & 0.5 & 7.1 & 0.6 \\
          & 70 & 5.9 & 0.7 & 6.1 & 0.9 & 6.1 & 0.4 \\
          & 50 & 5.5 & 0.6 & 5.7 & 1.0 & 5.7 & 0.5 \\
HD 199288 & 99 & 10.2 & 1.4 & 10.7 & 1.4 & 11.4 & 1.6 \\
          & 90 & 9.2 & 1.1 & 9.6 & 1.2 & 10.2 & 1.4 \\
          & 70 & 7.5 & 1.0 & 7.8 & 0.9 & 8.2 & 1.2 \\
          & 50 & 6.8 & 1.2 & 6.9 & 1.1 & 7.3 & 1.2 \\
HD 199509 & 99 & 25.7 & 10.6 & 29.4 & 14.7 & 33.7 & 19.8 \\
          & 90 & 23.1 & 10.0 & 22.7 & 8.7 & 23.7 & 8.8 \\
          & 70 & 15.9 & 3.3 & 16.7 & 4.2 & 21.1 & 9.3 \\
          & 50 & 13.3 & 2.4 & 13.4 & 2.3 & 15.7 & 4.7 \\
HD 202560 & 99 & 10.6 & 1.0 & 11.4 & 1.2 & 12.4 & 1.5 \\
          & 90 & 9.7 & 0.7 & 10.0 & 0.7 & 11.0 & 1.2 \\
          & 70 & 8.2 & 0.5 & 8.5 & 0.4 & 8.9 & 0.3 \\
          & 50 & 6.9 & 0.7 & 7.2 & 0.4 & 7.5 & 0.6 \\
HD 204385 & 99 & 14.6 & 0.7 & 15.7 & 0.8 & 17.9 & 2.2 \\
          & 90 & 13.5 & 0.7 & 14.1 & 0.5 & 15.4 & 0.8 \\
          & 70 & 11.3 & 1.8 & 11.9 & 0.9 & 12.5 & 1.2 \\
          & 50 & 9.1 & 1.7 & 9.6 & 1.6 & 10.0 & 1.7 \\
HD 204961 & 99 & 14.0 & 4.7 & 14.9 & 1.8 & 16.7 & 2.2 \\
          & 90 & 12.8 & 1.4 & 13.2 & 1.5 & 14.2 & 1.3 \\
          & 70 & 11.4 & 0.9 & 11.8 & 0.9 & 12.7 & 1.1 \\
          & 50 & 10.3 & 0.6 & 10.7 & 0.6 & 11.2 & 0.8 \\
HD 207129 & 99 & 8.7 & 1.2 & 9.3 & 1.4 & 10.3 & 1.7 \\
          & 90 & 7.5 & 1.1 & 8.0 & 1.1 & 8.6 & 1.1 \\
          & 70 & 5.9 & 0.9 & 6.2 & 0.8 & 6.3 & 0.7 \\
          & 50 & 4.7 & 1.0 & 4.8 & 0.5 & 5.0 & 0.7 \\
HD 208487 & 99 & 11.1 & 1.2 & 12.1 & 1.6 & 13.9 & 2.2 \\
          & 90 & 10.0 & 0.8 & 11.0 & 1.9 & 11.4 & 1.0 \\
          & 70 & 9.2 & 0.6 & 9.6 & 0.6 & 10.2 & 0.7 \\
          & 50 & 8.6 & 0.8 & 9.0 & 1.0 & 9.1 & 0.6 \\
HD 208998 & 99 & 18.9 & 1.9 & 20.9 & 5.0 & 24.4 & 11.3 \\
          & 90 & 17.4 & 1.2 & 18.3 & 1.6 & 20.2 & 3.0 \\
          & 70 & 16.4 & 1.6 & 17.1 & 2.5 & 18.1 & 3.0 \\
          & 50 & 15.2 & 1.3 & 15.5 & 1.7 & 16.1 & 1.4 \\
HD 209653 & 99 & 10.6 & 0.9 & 12.7 & 6.0 & 12.9 & 0.8 \\
          & 90 & 9.8 & 0.7 & 10.2 & 1.0 & 11.4 & 1.8 \\
          & 70 & 8.8 & 1.1 & 9.1 & 1.3 & 9.7 & 0.7 \\
          & 50 & 8.3 & 1.2 & 8.5 & 1.5 & 9.0 & 1.4 \\
HD 210918 & 99 & 13.6 & 11.1 & 14.5 & 12.3 & 15.9 & 9.7 \\
          & 90 & 13.5 & 11.8 & 13.0 & 9.5 & 14.6 & 11.1 \\
          & 70 & 6.9 & 3.7 & 7.5 & 4.3 & 7.7 & 4.3 \\
          & 50 & 4.6 & 3.0 & 4.7 & 3.1 & 5.0 & 3.3 \\
HD 211317 & 99 & 10.4 & 3.7 & 10.8 & 2.6 & 11.7 & 1.8 \\
          & 90 & 8.6 & 1.4 & 9.0 & 1.1 & 9.8 & 1.1 \\
          & 70 & 7.3 & 0.8 & 7.7 & 0.7 & 8.1 & 1.2 \\
          & 50 & 6.4 & 0.9 & 6.7 & 0.8 & 7.1 & 1.0 \\
HD 212168 & 99 & 10.6 & 0.7 & 11.7 & 0.7 & 12.9 & 1.0 \\
          & 90 & 9.7 & 0.6 & 10.2 & 0.3 & 10.9 & 0.5 \\
          & 70 & 8.4 & 0.5 & 9.0 & 1.0 & 9.6 & 1.4 \\
          & 50 & 7.6 & 0.5 & 7.9 & 0.6 & 8.3 & 0.8 \\
HD 212708 & 99 & 13.1 & 4.3 & 15.6 & 6.0 & 21.9 & 13.0 \\
          & 90 & 12.4 & 5.6 & 11.5 & 3.8 & 12.6 & 4.5 \\
          & 70 & 9.6 & 2.7 & 11.7 & 4.1 & 12.7 & 5.7 \\
          & 50 & 7.6 & 1.3 & 8.1 & 1.4 & 8.6 & 1.6 \\
HD 213240 & 99 & 16.5 & 7.8 & 19.1 & 9.3 & 21.2 & 9.6 \\
          & 90 & 12.5 & 3.0 & 14.1 & 4.4 & 15.4 & 4.7 \\
          & 70 & 11.3 & 3.3 & 13.5 & 5.1 & 14.7 & 6.2 \\
          & 50 & 9.1 & 1.3 & 10.0 & 2.0 & 10.8 & 2.4 \\
HD 214953 & 99 & 19.8 & 17.1 & 18.7 & 13.1 & 18.2 & 8.6 \\
          & 90 & 11.8 & 3.1 & 14.0 & 3.7 & 13.1 & 5.4 \\
          & 70 & 9.0 & 6.5 & 8.7 & 4.2 & 8.8 & 4.3 \\
          & 50 & 7.3 & 3.3 & 7.4 & 3.4 & 7.7 & 3.5 \\
HD 216435 & 99 & 8.5 & 0.5 & 8.9 & 0.7 & 10.1 & 2.1 \\
          & 90 & 8.3 & 0.5 & 8.4 & 0.5 & 8.7 & 0.6 \\
          & 70 & 7.7 & 0.7 & 7.7 & 0.9 & 8.2 & 1.0 \\
          & 50 & 6.9 & 0.6 & 7.0 & 0.7 & 7.3 & 0.8 \\
HD 216437 & 99 & 8.4 & 0.9 & 9.0 & 1.3 & 10.4 & 3.1 \\
          & 90 & 7.5 & 0.6 & 8.0 & 0.6 & 8.4 & 0.8 \\
          & 70 & 6.7 & 0.8 & 7.2 & 1.8 & 7.3 & 1.0 \\
          & 50 & 6.0 & 0.6 & 6.3 & 0.5 & 6.6 & 0.6 \\
HD 217958 & 99 & 32.3 & 7.3 & 36.0 & 9.1 & 43.8 & 14.7 \\
          & 90 & 30.7 & 11.7 & 30.0 & 6.5 & 35.3 & 9.8 \\
          & 70 & 27.6 & 7.8 & 26.8 & 6.8 & 31.9 & 10.0 \\
          & 50 & 23.3 & 5.4 & 23.4 & 4.9 & 26.2 & 6.8 \\
HD 217987 & 99 & 22.0 & 3.8 & 24.0 & 4.9 & 27.5 & 6.2 \\
          & 90 & 20.1 & 2.5 & 21.0 & 3.0 & 22.8 & 3.4 \\
          & 70 & 18.6 & 3.2 & 18.2 & 1.3 & 19.6 & 1.6 \\
          & 50 & 16.6 & 1.5 & 17.0 & 1.5 & 18.1 & 2.4 \\
HD 219077 & 99 & 13.5 & 8.2 & 14.0 & 7.5 & 15.5 & 8.8 \\
          & 90 & 9.5 & 2.8 & 10.6 & 3.4 & 11.0 & 3.9 \\
          & 70 & 6.3 & 2.2 & 7.1 & 1.7 & 7.1 & 2.4 \\
          & 50 & 4.0 & 2.4 & 4.0 & 2.5 & 4.2 & 2.6 \\
HD 221420 & 99 & 12.3 & 8.2 & 14.2 & 12.1 & 14.2 & 9.5 \\
          & 90 & 8.2 & 1.7 & 9.9 & 4.1 & 10.3 & 4.1 \\
          & 70 & 5.2 & 2.2 & 6.1 & 1.5 & 6.6 & 4.1 \\
          & 50 & 4.4 & 2.2 & 4.7 & 2.2 & 4.9 & 2.3 \\
HD 223171 & 99 & 9.1 & 1.5 & 10.3 & 2.9 & 12.8 & 8.5 \\
          & 90 & 8.1 & 0.8 & 8.6 & 1.0 & 9.5 & 1.5 \\
          & 70 & 7.2 & 1.8 & 7.4 & 0.7 & 7.5 & 0.9 \\
          & 50 & 5.9 & 1.0 & 6.5 & 1.6 & 6.9 & 2.2 \\
\enddata
\end{deluxetable}


\begin{deluxetable}{lr@{$\pm$}lr@{$\pm$}lr@{$\pm$}lr@{$\pm$}lr@{$\pm$}lr@{$\pm$}
lr@{$\pm$}ll}
\rotate
\tabletypesize{\scriptsize}
\tablecolumns{9}
\tablewidth{0pt}
\tablecaption{Substellar Companions From This Sample\tablenotemark{a} }
\tablehead{
\colhead{Planet} & \multicolumn{2}{c}{Period} & \multicolumn{2}{c}{$T_0$}
&
\multicolumn{2}{c}{$e$} & \multicolumn{2}{c}{$\omega$} &
\multicolumn{2}{c}{K } & \multicolumn{2}{c}{M sin $i$ } &
\multicolumn{2}{c}{$a$ } & \colhead{Discovery Ref.} \\
\colhead{} & \multicolumn{2}{c}{(days)} & \multicolumn{2}{c}{(JD-2400000)}
&
\multicolumn{2}{c}{} &
\multicolumn{2}{c}{(degrees)} & \multicolumn{2}{c}{(\ms)} &
\multicolumn{2}{c}{(\Mjup)} & \multicolumn{2}{c}{(AU)} & \colhead{}
 }
\startdata
\label{planets}   
HD 142 b & 350.0 & 3.6 & 51960 & 43 & 0.3 & 0.18 & 303 & \nodata &
34 & 4.7 & 1.3 & 0.32 & 1.045 & 0.061 & \citet{tinney02} \\
HD 2039 b & 1166 & 11 & 54405 & 9 & 0.574 & 0.048 & 344 & 5 &
72.1 & 4.3 & 3.77 & 0.45 & 2.41 & 0.13 & \citet{tinney03} \\
HD 4308 b & 15.609 & 0.007 & 50108.5 & 1.9 & 0.27 & 0.12 & 210 & 21 &
3.6 & 0.3 & 13.0 & 1.4 & 0.118 & 0.009 & \citet{udry06} \\
HD 16417 b & 17.24 & 0.01 & 50099.7 & 3.3 & 0.20 & 0.09 & 77 & 26 &
5.0 & 0.4 & 0.067 & 0.009 & 0.14 & 0.01 & \citet{16417paper} \\
HD 20782 b & 585.86 & 0.03 & 51687 & 2.5 & 0.93 & 0.03 & 147 & 3 &
120 & 12 & 1.8 & 0.3 & 0.14 & 0.01 & \citet{jones06} \\
HD 23127 b & 1211 & 16 & 53649 & 31 & 0.396 & 0.090 & 183 & 10 &
27.4 & 1.9 & 1.52 & 0.13 & 2.39 & 0.08 & \citet{otoole07} \\
HD 27442 b & 428 & 1.1 & 50840 & 55 & 0.06 & 0.04 & 216 & \nodata &
32.0 & 1.4 & 1.5 & 0.5 & 1.27 & 0.07 & \citet{butler01} \\
HD 39091 b & 2086 & 3 & 50036 & 3 & 0.638 & 0.005 & 330.6 & 0.7 &
194.4 & 1.2 & 10.02 & 0.20 & 3.31 & 0.03 & \citet{jones02a} \\
HD 70642 b & 2167 & 21 & 51853 & 177 & 0.068 & 0.039 & 295 & 29 &
27.8 & 1.1 & 1.82 & 0.11 & 3.33 & 0.05 & \citet{carter03} \\
HD 75289 b & 3.50927 & 0.000064 & 50830.3 & 0.475 & 0.03 & 0.03 & 141 & 
\nodata & 55 & 1.8 & 0.46 & 0.04 & 0.048 & 0.003 & \citet{udry00} \\
HD 76700 b & 3.9710 & 0.0002 & 51213.3 & 0.7 & 0.10 & 0.08 & 30 & \nodata &
28 & 1.7 & 0.23 & 0.03 & 0.0511 & 0.0030 & \citet{tinney03} \\
HD 102117 b & 20.813 & 0.006 & 50942 & 3 & 0.12 & 0.08 & 279 & \nodata &
12.0 & 1.0 & 0.17 & 0.02 & 0.1532 & 0.0088 & \citet{tinney05} \\
HD 117618 b & 25.83 & 0.02 & 50832 & 2 & 0.4 & 0.17 & 250 & 19 &
13 & 2.2 & 0.18 & 0.05 & 0.175 & 0.010 & \citet{tinney05} \\
HD 134987 b & 258.19 & 0.07 & 50071.0 & 0.8 & 0.233 & 0.002 & 352.7 & 0.5 &
49.5 & 0.2 & 1.59 & 0.02 & 0.81 & 0.02 & \citet{jones10} \\
HD 160691 b & 643.25 & 0.90 & 52366 & 13 & 0.13 & 0.017 & 22 & 7 &
37.8 & 0.4 & 1.7 & 0.13 & 1.497 & \nodata & \citet{butler01} \\
HD 160691 d & 9.6386 & 0.0015 & 52991.1 & 0.4 & 0.17 & 0.04 & 210 & 13 &
3.1 & 0.13 & 0.035 & 0.003 & 0.0909 & \nodata & \citet{santos04} \\
HD 160691 e & 310.55 & 0.83 & 52708.7 & 8.3 & 0.07 & 0.01 & 189.6 & 9.4 &
14.9 & 0.6 & 0.54 & 0.04 & 0.921 & \nodata & \citet{pepe07} \\
HD 164427 b & 108.55 & 0.04 & 51724.6 & 0.2 & 0.55 & 0.02 & 356.9 & 0.5 &
2229 & 77 & 46.4 & 3.4 & 0.46 & 0.05 & \citet{tinney01} \\
HD 179949 b & 3.09251 & 0.00003 & 51002.4 & 0.4 & 0.02 & 0.015 & 192 & 
\nodata & 113 & 1.8 & 0.90 & 0.07 & 0.0443 & 0.0026 & \citet{tinney01} \\
HD 187085 b & 1065 & 19 & 51392 & 325 & 0.047 & 0.090 & 261 & 114 &
15.3 & 1.4 & 0.88 & 0.13 & 2.19 & 0.08 & \citet{jones06} \\
HD 196050 b & 1398 & 15 & 54973 & 55 & 0.181 & 0.030 & 174 & 11 &
49.4 & 1.4 & 3.02 & 0.22 & 2.60 & 0.07 & \citet{jones02b} \\
HD 208487 b & 130.1 & 0.51 & 51000 & 16 & 0.2 & 0.16 & 113 & \nodata &
20 & 3.6 & 0.5 & 0.13 & 0.524 & 0.030 & \citet{tinney05} \\
HD 213240 b & 883 & 7.6 & 51500 & 13 & 0.42 & 0.02 & 201 & 3 &
97 & 2 & 4.5 & 0.34 & 1.92 & 0.11 & \citet{santos01} \\
HD 216435 b & 1339 & 16 & 50632 & 158 & 0.069 & 0.062 & 41 & 42 &
19.8 & 1.1 & 1.28 & 0.12 & 2.60 & 0.06 & \citet{jones03} \\
HD 216437 b & 1355 & 7 & 51942 & 18 & 0.357 & 0.025 & 61 & 5 &
39.0 & 1.1 & 2.22 & 0.08 & 2.54 & 0.03 & \citet{jones02b} \\
\hline
 \multicolumn{5}{c}{Jupiter analogs} \\
HD 134987 c & 5000 & 400 & 51100 & 600 & 0.12 & 0.02 & 195 & 48 &
9.3 & 0.3 & 0.82 & 0.03 & 5.8 & 0.5 & \citet{jones10} \\
HD 160691 c & 4163 & 99 & 52513 & 62 & 0.029 & 0.024 & 23 & 48 &
23.2 & 0.5 & 2.00 & 0.10 & 5.3 & 0.1 & \citet{mccarthy04} \\
GJ 832 b & 3380 & 189 & 54618 & 374 & 0.157 & 0.015 & 305 & 40 & 
14.4 & 2.0 & 0.62 & 0.13 & 3.38 & 0.40 & \citet{bailey09} \\
\enddata
\tablenotetext{a}{Those 123 stars with $N>30$ and $T_{obs}>$8~yr.}
\end{deluxetable}

\begin{deluxetable}{llll}
\tabletypesize{\scriptsize}
\tablecolumns{4}
\tablewidth{0pt}
\tablecaption{Jupiter-Analog Upper Limits from the AAPS Sample }
\tablehead{
\colhead{Velocity Amplitude} & \multicolumn{3}{c}{Upper Limit} \\
\colhead{(\ms)} & \multicolumn{3}{c}{percent}\\
\colhead{} & \colhead{e=0.0} & \colhead{e=0.1} & \colhead{e=0.2} 
 }
\startdata
\label{upperlimits}
$K>50$ & 11.6$\pm$1.1 & 12.3$\pm$1.4 & 14.6$\pm$1.5 \\ 
$K>40$ & 12.6$\pm$1.1 & 13.6$\pm$1.4 & 16.2$\pm$1.5 \\
$K>30$ & 14.4$\pm$1.2 & 15.4$\pm$1.4 & 18.6$\pm$1.5 \\
$K>20$ & 18.6$\pm$1.1 & 20.7$\pm$1.5 & 23.8$\pm$1.6 \\
$K>10$ & 37.2$\pm$1.1 & 44.8$\pm$1.4 & 48.8$\pm$1.5 \\
\enddata
\end{deluxetable}

\end{document}